\documentclass[5p,sort&compress]{elsarticle}

\usepackage{latexsym}
\usepackage{amsfonts,amsmath,amssymb}
\usepackage{url}
\usepackage[utf8]{inputenc}
\usepackage[T1]{fontenc}
\usepackage{lmodern}
\usepackage{fancyref}
\usepackage{url}
\usepackage{mathtools}
\usepackage{booktabs}
\usepackage[flushleft]{threeparttable}
\usepackage[english]{babel}
\usepackage[shortcuts]{extdash}
\usepackage{listings}
\usepackage[inline]{enumitem}
\usepackage[labelformat=simple]{subcaption}
\usepackage{algorithm}
\usepackage{algpseudocode}
\usepackage[draft]{todonotes}
\usepackage{multirow}
\usepackage{footnote}
\makesavenoteenv{tabular}

\newcommand{\nodewatcher}{\textit{nodewatcher}}
\newcommand{\wlanslovenija}{\textit{wlan slovenija}}

\begin{document}

\begin{frontmatter}
\title{\nodewatcher{}: A Substrate for Growing Your own Community Network}

\author[fri,wlansi]{Jernej Kos\corref{cor}}
\ead{jernej.kos@fri.uni-lj.si}

\author[berkeley,wlansi]{Mitar Milutinović}
\ead{mitar.nodewatcher@tnode.com}

\author[fri,wlansi]{Luka Čehovin}
\ead{luka.cehovin@fri.uni-lj.si}

\cortext[cor]{Corresponding author}

\address[fri]{University of Ljubljana, Faculty of Computer and Information Science, Ljubljana, Slovenia}

\address[berkeley]{University of California, Berkeley, USA}

\address[wlansi]{\wlanslovenija{}, Open wireless network of Slovenia, \url{https://wlan-si.net}}

\begin{abstract}
Community networks differ from regular networks by their organic growth patterns -- there is no central planning body that would decide how the network is built.
Instead, the network grows in a bottom-up fashion as more people express interest in participating in the community and connect with their neighbours.
People who participate in community networks are usually volunteers with limited free time.
Due to these factors, making the management of community networks simpler and easier for all participants is the key component in boosting their growth.
Specifics of individual networks often force communities to develop their own sets of tools and best practices which are hard to share and do not interoperate well with others.
We propose a new general community network management platform \nodewatcher{} that is built around the core principle of modularity and extensibility, making it suitable for reuse by different community networks.
Devices are configured using platform-independent configuration which \nodewatcher{} can transform into deployable firmware images, eliminating any manual device configuration, reducing errors, and enabling participation of novice maintainers.
An embedded monitoring system enables live overview and validation of the whole community network.
We show how the system successfully operates in an actual community wireless network, \wlanslovenija{}.
\end{abstract}

\end{frontmatter}

\section{Introduction}

Community (wireless) networks~\cite{Bruno_2005,Frangoudis_2011} provide independent, community-owned network infrastructure for user communication and data exchange.
They are mostly built using standard wireless (IEEE802.11) infrastructure~\cite{Akyildiz_2005}, by laying own optical fiber and, more recently, by the use of free-space optical systems~\cite{Mustafa_2013}.
They use, reuse and repurpose existing communication technologies, like inexpensive off-the-shelf WiFi routers, to form a widespread network.
These networks can cover everything from local neighbourhoods~\cite{RedHook_2013}, to cities~\cite{AWMN} and countries~\cite{wlanslovenija_2009, guifi_2003, Funkfeuer_2003, Freifunk_2003}.
Their common aim is to empower people with new ways of communication and access to the wider public networks like the Internet.
Motivations for such networks are diverse and multiplex, but often networks are formed out of necessity~\cite{WNDW_2013}.
Even in developed countries many rural areas are underserved with Internet connectivity infrastructure built and offered by traditional Internet service providers, population density being too low for investments to be profitable enough, leaving to people themselves to build needed infrastructure to connect to the Internet.

\begin{figure}[t]
  \centering
  \includegraphics[width=\columnwidth]{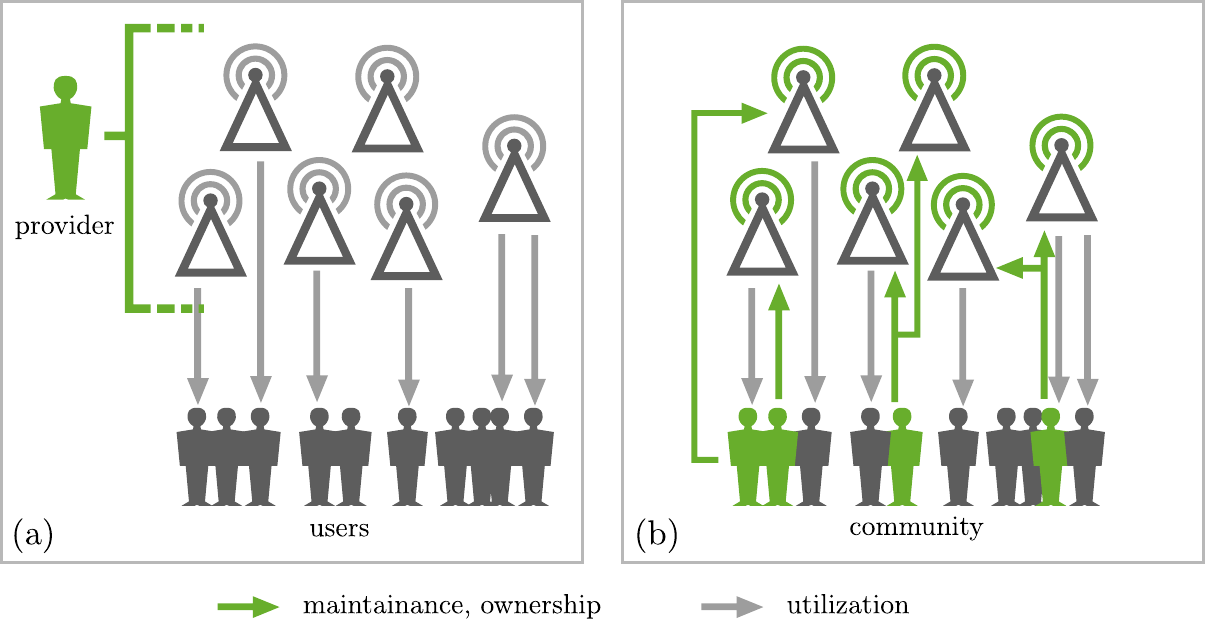}
  \caption{Illustration of conceptual difference between proprietary networks (a) and community networks (b).}
  \label{fig:community}
\end{figure}

The common property of all community networks is that they grow organically~-- there is no central planning body that would decide how and where the network is built, as is usually the case with proprietary networks.
Instead, the network grows in a bottom-up fashion as more people express interest in participating in the community and connect with their neighbours as illustrated in Figure~\ref{fig:community}.
When a new person or a new group decides to expand the community network, they can do this themselves by deploying new access points, routers, antennas, and other devices, connecting them to the rest of the network, effectively growing it.
Often only very local coordination, with direct neighbours, is needed.
Through this process, both the community and the network grows further, enabling even more people to participate in the network in the future.
Because of this bottom-up growth pattern and community involvement, management of such networks poses some unique challenges:

\begin{itemize}
\item In most community networks, people who maintain the network infrastructure are volunteers with limited free time that they can spend on network management.
This makes efficient management very important for network growth.
Besides efficiency, an important factor in increasing growth is also the accessibility of network management functions to users who do not have deep technical knowledge of computer networks and especially mesh networks.

\item There are many repetitive tasks in community network operation, mostly related to configuration, deployment and monitoring of network equipment.
Without a suitable overall management system, all these steps (flashing\footnote{Flashing is the process of reprogramming the wireless network device with an unofficial firmware image, commonly Linux based.} and configuring devices, allocating resources, diagnosing problems) need to be done manually which is both time-consuming and prone to errors.

\item In addition to technical issues related to device deployment, there is also the need for community coordination so that people know what is going on in the network and can familiarize themselves better with its operation and what others are doing.
In general it is not possible to anticipate how, when, and where a new participant will start participating by deploying one of the network's nodes\footnote{Meaning of a ``network node'' is understood differently in different communities. For the purpose of this paper it is used as one routing network device, and as a basic unit of participation by participants.}.

\item For volunteers it is important to have feedback on how they themselves are contributing to the network.
Is the node that they are maintaining highly used and crucial for the part of the network and users?
Volunteers are often motivated by the value they are contributing to the network as a whole, but even when their node is not highly used at the moment, having an insight into operation of the node is important for them to understand that their contribution is something tangible and real, especially for less technically skilled volunteers who might otherwise perceive the node as a black box.
\end{itemize}

Various solutions have appeared in an attempt to address these challenges. Each community developed its own model of operation with its own accompanying set of tools.
The problem with this approach is that work is being duplicated between communities and that these various solutions are mostly not interoperable between each other.
So why would new communities not reuse existing tools?
The problem lies in the fact that these tools have not been built to be customized to the needs and operation of individual communities.
Each community has slight differences in their vision and operation philosophy, technological stack they are using and technical knowledge they have, or the local environment they are in, and this requires customization on several fronts, to only name a few:
\begin{enumerate*}[label=\itshape\alph*\upshape)]
\item different routing protocols may be used;
\item used WiFi equipment and its operating systems can vary widely between networks;
\item some communities use VPN tunnels to establish certain long-range links using different VPN protocols;
\item network topologies may differ among communities, some use central clusters of nodes as gateways to peer with other networks, some use a more distributed topology;
\item some communities attach various sensors to nodes and would like to monitor their outputs through time;
\item networks can decide to use a captive portal and/or choose to require users to pay to use the network;
\item there may be differences in operation due to local regulations.
\end{enumerate*}

In order to address this, there are at least two approaches that can be taken.
The first one is \textit{the large common base approach} which tries to create a common system that addresses the needs of as many of communities as possible by providing a large feature set and a large configuration schema (database model) encompassing all possible scenarios.
This approach has been tried as part of an interoperability effort~\cite{interop_2010, cnml_2007} established between community networks, to come up with a common schema upon which node databases\footnote{``Node databases'' is a generic term communities use for their network management needs. They can be everything from a simple wiki page with a list of nodes to a fully-fledged and specialized software solutions.} could be built.
The problem is that it is hard to come up with a one-fits-all solution and large monolithic schemas can quickly become unmanageable.
Moreover, it requires sustained effort from participating community networks first to establish the standard and then to keep it updated as technologies and practices evolve.
Because community networks are mostly volunteer run, such sustained participation is unattainable for many community networks.
Additionally, it is practically impossible to involve all community networks in this process and there are always new networks with new differences.
This process favours large and established network communities.

The other way is \textit{the extensible core approach} where the aim is to create a minimal core with highly modular and extensible design so that community networks can tailor it to their needs, whatever they might be.
This is the approach that we are taking with the \nodewatcher{} v3 platform.
We make the following novel contributions:
\begin{itemize}
\item A modular open platform that may be easily tailored to the needs of any community network.
\item An extensible per-node firmware image generation system that enables generation of pre-configured images for specific nodes in order to eliminate any manual configuration requirement on the nodes.
\item An extensible monitoring system with a scalable time-series data storage backend enabling large-scale collection of status and other telemetry data while supporting interactive visualizations.
\item User interface designed and structured so that it is suitable both for novice and expert users, tailored for collaboration and coordination of volunteers.
\end{itemize}

The rest of this paper is organized as follows.
Section~\ref{sec:related-work} presents related work done in the area of community network management tools.
Section~\ref{sec:platform} presents the design and functioning of the \nodewatcher{} platform.
Section~\ref{sec:evaluation} shows the results of platform evaluation in the \wlanslovenija{} community wireless network.
Section~\ref{sec:conclusion} presents conclusions and ideas for future work.

\section{Related Work}
\label{sec:related-work}

A lot of research has been done into individual wireless mesh network building blocks. Among them are routing~\cite{Murray_2010,Neumann_2012,Neumann_2013}, security~\cite{Siddiqui_2007}, and analyses of topologies, performance, mobility~\cite{Vega_2012,Zakrzewska_2008,Braem_2014,Cerda_2013,Maccari_2015}.
But research into community network management solutions and best practices still remains scarce.
This is why most of the related work in this area comes from the individual community networks which have each developed its own solutions, practices and philosophy.

In this section we survey the most visible network management solutions, traditional ones and specialized solutions developed by community networks worldwide.
We compare them to the approach taken by \nodewatcher{} platform.

\subsection{Traditional Network Management}

There are many existing network management systems, suited for more traditional (non-community) networks~\cite{Cacti_2004,Nagios_1999,SmokePing_2001,Zabbix_2004,Puppet_2005,Salt_2011}.
By their management function, they can be segmented into two major classes as follows:
\begin{description}
\item[Monitoring systems] enable the operators to remotely monitor a set of devices to see whether they are reachable and to get some insights into their operation.
Some of these systems can generate events and notify the operators when errors are detected, while others can only visualize the data without interpreting it.

\item[Configuration management systems] enable the operators to maintain a central repository of device configurations that can be used to provision devices.
In these systems, configuration is mostly input using a domain-specific or a scripting language and deployment is done using remote agents that interpret configuration scripts and apply configurations.
\end{description}

While these systems can be used in community networks, they are not tailored to their specific needs.
First, some of the monitoring and configuration management systems require agents which consume too much resources to run on simple off-the-shelf network equipment commonly in use in community networks.
They might be suitable to monitor some better-equipped network nodes, but would require different systems for managing other parts of the network, which increases the work load of volunteer participants.
Additionally, these systems are independent solutions, requiring manual integration which needs to be performed by each community.
This increases the chance that each community will pick different solutions and end up with systems which are not interoperable with each other.
Configuration management systems are not suitable for embedded systems, or if they are, they target devices of a particular manufacturer, and not a diverse range of often customized devices found in community networks.

But most importantly, since they have not been designed with community networks in mind, none of these systems provide community coordination capabilities.
They are highly technical to setup and use, which makes it harder to properly configure by communities without highly skilled members.
They are designed for general networks and do not encode experiences and particularities of community networks in their source code, which would help new community networks to start with reasonable and tested defaults and configurations.
Reading information from monitoring systems is designed for trained network operators and entering configuration into management systems requires good understanding of computer networks terminology and operation.
They are often developed as an independent codebase which makes it much harder to integrate with other open source solutions used by a community network, like community management solutions, and they lack one unified interface, which further confuses novice users.

\subsection{Node Databases}

\newcommand{\push}{$\uparrow$}
\newcommand{\pull}{$\downarrow$}

\newcommand{\no}{$\times$}
\newcommand{\yes}{$\checkmark$}
\newcommand{\maybe}{$\circ$}

\newcommand{\static}{S}
\newcommand{\dynamic}{D}

\begin{table*}[t!]
\caption{\label{tbl:comparision}Feature comparison of existing node database systems as of May 2015. In the table we show the developer group of a project, if it is actively developed (A) and when it was first officially mentioned and deployed (Y). In terms of project functionalities we list network monitoring (NM), federation support (FDR), configuration generation (CF), firmware generator (FW), application program interface (API), resource allocation (RA), node authentication (NA), link planning (LP), topology visualization (TP), and map visualization (MP). We also show which projects can be easily extended in terms of user interface (UI), application program interface (API), network monitoring (NM), support of platforms (PL), schema (SC), and routing protocols (RP). In terms of developer accessibility we highlight the documentation status (DOC), programming language (LAN), web framework (WF), and code license (LIC).}

\bgroup
\def\arraystretch{1.5}

\centering
\renewcommand{\tabcolsep}{2pt}
\scriptsize{
\begin{tabular}{|p{9.5em}|c|p{3em}|c|c|c|c|c|c|c|c|c|c|c|c|c|c|c|c|c|p{4em}|c|c|}
\hline
\multirow{2}{*}{Project (Developers)} & \multicolumn{2}{c|}{General}                                   & \multicolumn{10}{c|}{Features}                                                                                                                                 & \multicolumn{6}{c|}{Modularity/Extensibility}                                          & \multicolumn{4}{c|}{Developer Accessibility}                                            \\ \cline{2-23} 
      & A & Y & NM & FDR & CF & FW & API     & RA & LP & NA & TP & MP     & UI & API & NM & PL & SC  & RP & DOC & LAN    &  WF & LIC        \\ \hline
Guifi.net frw. \newline (Guifi.net)            & \yes            & 2004     \newline 2004     & \maybe               & \no         & \yes                    & \no               & \yes     & \yes                & \yes          & \no         & \static      & \yes     & \no            & \yes & \no               & \no             & \no     & \no                & \yes           & PHP                    & Drupal       & GPLv2            \\ \hline
AWMN WiND      \newline (AWMN)                 & \yes            & 2005    \newline 2005    & \no              & \no         & \no                     & \no                & \yes     & \yes                & \yes          & \no         & \static      & \yes     & \no            & \no  & \no                & \no              & \no      & \no               & \no                & PHP                    & custom       & AGPLv3              \\ \hline
FFM/CNDB       \newline (FunkFeuer)             & \yes            & 2012     \newline 2015        & \no             & \no          & \no                    & \no               & \yes       & \yes                  & \no            & \no         & \no     & \yes       & \yes             & \yes   & \no                 & \no              & \yes       & \no              & \yes                 & PY                 & custom      & BSD              \\ \hline
Nodeshot       \newline (Ninux)                & \yes            & 2011     \newline 2011     & \no              & \yes        & \no                     & \no                & \yes     & \no                 & \yes          & \yes       & \dynamic      & \yes     & \no            & \yes & \no                & \no              & \maybe & \yes          & \yes               & PY                 & Django       & GPLv3          \\ \hline
Netmon         \newline (Freifunk)             & \yes            & 2009     \newline 2009    & \push, \pull           & \no        & \no                    & \no               & \yes     & \yes                  & \no            & \no         & \dynamic     & \yes     & \no             & \yes   & \no                 & \no              & \no     & \no          & \no                 & PHP                    & custom            & GPLv3              \\ \hline
LibreMap       \newline (Altermundi)       & \yes            & 2013     \newline 2013        & \push               & \yes        & \no                     & \no                & \yes       & \no                 & \no           & \no         & \dynamic     & \yes     & \no             & \yes   & \yes                 & \no               & \yes       & \yes         & \no                 & JS                      & custom            & GPLv3          \\ \hline
kalua          \newline (weimarnetz.de)        & \yes            & 2001     \newline 2001     & \push           & \no         & \yes                    & \yes               & \no      & \yes                & \no           & \yes       & \dynamic      & \yes     & \no            & \no  & \no                & \no              & \no      & \yes       & \no                 & Sh, PHP                  & custom          & GPLv2            \\ \hline
meshviewer     \newline (Freifunk L\"{u}beck)  & \yes            & 2015     \newline 2015     & \push, \pull    & \no         & \no                     & \no                & \yes     & \no                 & \no           & \no         & \dynamic      & \yes     & \no            & \no  & \no                & \no              & \no      & \yes             & \no                 & JS             & custom            & AGPLv3         \\ \hline
K-Net          \newline (DTU Students)         & \yes            & 1996     \newline 1996     & \pull           & \no         & \no                     & \no                & \yes     & \maybe             & \yes          & \no         & \dynamic      & \yes     & \yes           & \yes & \no                & \no              & \no      & \no              & \no                 & C, C++, PY, HS & Django       & none \\ \hline
Geronimo       \newline (Opennet Initiative)   & \yes            & 2011     \newline 2011     & \no             & \no         & \no                      & \no                 & \yes     & \no                & \no           & \no         & \dynamic     & \yes     & \no           & \yes & \no                 & \no             & \maybe     & \no                & \yes               & PHP, PY           & Django       & none              \\ \hline
Ondataservice  \newline (Opennet Initiative)   & \no            & 2009     \newline 2009     & \push             & \yes          & \no                      & \no                 & \yes     & \no                & \no           & \no         & \dynamic      & \no      & \no            & \no & \no                 & \yes             & \no     & \no              & \no               & C, PHP                 & custom            & BSD              \\ \hline
NCD            \newline (Routek/qMp/Guifi)     & \yes            & 2014     \newline 2015     & \pull           & \no         & \no                     & \no                & \no      & \no                 & \no           & \no   & \dynamic      & \no & \yes           & \no   & \yes               & \no              & \yes     & \yes              & \no           & Lua, JS                & D3.js           & GPLv3         \\ \hline
manman         \newline (Funkfeuer Graz)       & \no             & 2006     \newline 2006     & \pull             & \no         & \no                     & \no                & \maybe & \yes                & \no            & \no        & \no    & \yes     & \yes           & \no  & \yes               & \no              & \yes     & \no                & \no                 & RB, PL              & Rails        & none              \\ \hline
nodewatcher v2 \newline (wlan slovenija)       & \no             & 2009     \newline 2009     & \pull             & \no         & \yes                    & \yes               & \no      & \yes                & \no           & \no        & \dynamic      & \yes    & \no            & \no  & \no                & \no              & \no       & \no                & \no                & PY                 & Django       & AGPLv3         \\ \hline
nodewatcher v3 \newline (wlan slovenija)       & \yes            & 2012     \newline 2015     & \push, \pull      & \maybe        & \yes                    & \yes               & \yes     & \yes                & \no           & \yes       & \dynamic      & \yes     & \yes           & \yes & \yes               & \yes             & \yes     & \yes              & \yes               & PY, C              & Django       & AGPLv3         \\ \hline
\end{tabular}
}

\egroup
\end{table*}

Many wireless mesh network communities have quickly recognized the need for having a central system that would be able to manage the growing number of wireless mesh nodes.
We present a comprehensive overview of properties of existing solutions in Table~\ref{tbl:comparision}. The data was obtained by studying the public information available about the projects (e.g. source code) and verified by reaching out to the members of the corresponding communities.
We list several properties of community network management solutions, starting with the name of the project and the community that forms its core development group.
We show whether the project is being actively developed (A) and when was it first officially mentioned and deployed in an actual network (Y).
Properties are then organized into three major areas, namely  features (functionality that the system has), modularity/extensibility (can the individual functionalities be extended without modifying the core) and developer accessibility.
For most of the features, we list whether the described system implements the feature fully (\yes), only partially (\maybe) or not at all (\no).
In case of network monitoring (NM), we also list whether the system supports pulling data from nodes (\pull) and if nodes can push data to the system (\push).
For topology visualization (TP), we mark whether the system renders the topology just based on configuration (\static) or it can also display live topology as it changes based on some routing protocol (\dynamic).

One of the oldest and largest mesh networks are Guifi.net~\cite{Guifinode_2003,Vega_2012} and the Athens Wireless Metropolitan Network (AWMN)~\cite{AWMN_WIND_2002}.
As the node database solutions evolved with their respective networks, both are tailored to specific structures of their networks and its management structure.
Their codebase is monolithic, making it hard to extend with new features or customizations.
This applies to the schema (which lacks any kind of object-relational mapping~\cite{ONeil_2008} and is therefore hard to manage when developing extensions), frontend interface and core functionalities.
AWMN WIND is built on a custom web framework, further limiting the ease of adoption by a new community network.
Additionally, advanced network monitoring functionality is missing, requiring the use of external utilities, which causes duplication of configuration and is an error-prone process (see Section~\ref{sec:network-monitoring}).
The Guifi.net framework does support including network monitoring graphs, but it requires the use of external monitoring services to perform the actual monitoring.

Two more recent representatives of community network node databases are Nodeshot~\cite{Nodeshot_2012} and FFM/CNDB~\cite{Funkfeuer_2012}, developed by Ninux and FunkFeuer community networks, respectively.
Nodeshot is an extensible web application for management of community geographic data.
It focuses on mapping features with a more modular approach where various functionalities are extracted into modules.
The modularity makes extending the application easier, but in contrast to \nodewatcher{}, there is no common approach to managing schema extensions which makes the database models hard to extend in a modular fashion.
There is some monitoring support, but it is mostly limited to device discovery in an existing network and is not meant for long-term monitoring and diagnostics.

FFM/CNDB features an extensive database schema which models everything from devices to companies and people.
With its very detailed modelling of individual objects it is an example of the large common base approach.
Similar to Nodeshot, there is no common approach to managing schema extensions and there is also no monitoring support.
It uses a custom web framework with almost nonexistent documentation, having a steep learning curve before extending and adapting the system to another community network is possible.
While the Guifi.net framework does support generating device configuration, none of the existing solutions support generating functional and pre-configured firmware images that can be directly flashed to devices.
This would reduce the required administration burden and make the whole process easier for novices.
As a consequence, these solutions also do not support comparing running configuration with provisioned one and detecting misconfiguration and unwanted changes.
This makes detecting possible issues in advance impossible, and debugging issues once they occur slower.

In this section we have presented an overview of existing community network management solutions.
From these, we summarize a list of open problems:
(a) existing solutions mostly lack modularity and easy extensibility in several areas including user interface, network monitoring, schema, configuration/firmware generation and routing protocols;
(b) there is limited support for generating configuration, and no support for generating pre-configured firmware images;
(c) there is limited support for interactive display of time-series monitoring data.
The next section presents possible solutions to these problems.

\section{Platform}
\label{sec:platform}

The design and development of the \nodewatcher{} platform comes from the needs and evolution of the \wlanslovenija{} community wireless network.
In \wlanslovenija{} we had similar needs and issues as other community networks and to address them we iteratively developed our own node database system, tailored to our needs and practices.
Those were versions v1 and v2 of the \nodewatcher{} platform.
But as we started to collaborate with other community networks, we soon discovered many of more broader issues outlined in Section~\ref{sec:related-work}.
We learned that if we want our development efforts to benefit also other community networks and help grow new ones, a different approach is needed.
One where we do not focus development only on the needs of our own network, but think about other networks and their use of the platform from the very beginning.
As a consequence, the \nodewatcher{} v3 platform has been designed to be maximally extensible and reusable among different communities.
This section describes all the components of the platform, beginning with a quick high-level overview of the architecture.

\subsection{Overview}

\begin{figure}
  \centering
  \includegraphics[width=\columnwidth]{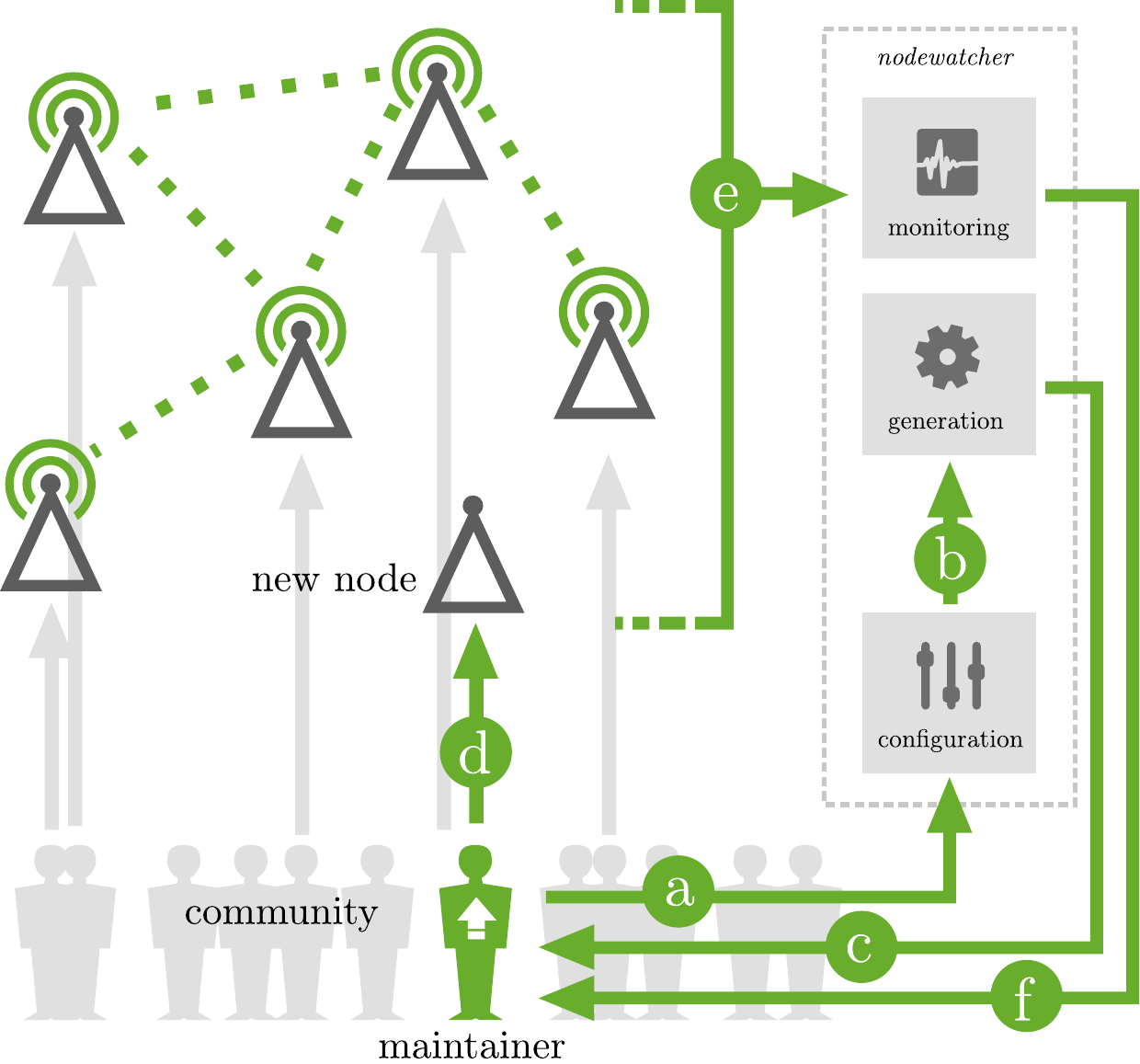}
  \caption{Illustration of the device management cycle in a \nodewatcher{}-supported community network. Traditionally the configuration (a), generation (b) and monitoring (e) steps are performed manually, while \nodewatcher{} enables automation of these tasks, freeing up resources inside the community. On the other hand, the deployment step (c) is still community-driven, allowing the network to automatically grow and adapt with it.}
  \label{fig:device-mgmt-cycle}
\end{figure}

The core idea behind \nodewatcher{} is the \textit{device management cycle}, illustrated by Figure~\ref{fig:device-mgmt-cycle}.
The management cycle is a model of how devices are managed in a community network.
First, members of the community (usually the node maintainer) decide that a node should be deployed or reconfigured.
This is handled in the \textit{configuration stage} (Figure~\ref{fig:device-mgmt-cycle}, step a), where devices are configured to be used at a specific location in the network.
At this stage, the device configuration does not depend on the hardware that will be later used to deploy the device, making it platform-independent (Section~\ref{sec:platform-independent-configuration}).
Configuration can be adjusted by members of the community using a dynamically generated web interface (Section~\ref{sec:form-generation}), reflecting the schema that is used for describing configuration.
Additionally in the course of configuration, \nodewatcher{} may automatically allocate the resources (e.g. IP addresses) needed for proper functioning of a device in the network, taking into account any defined allocation policies (Section~\ref{sec:resource-allocation}).
The next stage is the \textit{generation stage} (Figure~\ref{fig:device-mgmt-cycle}, step b), where platform-independent configuration is transformed to device- and platform-specific configuration. 
If the transformation is successful, a firmware image, suitable for the target device, may also be generated (Section~\ref{sec:firmware-generator}) in order to ease deployment.
In case of successful outcome of the generation process (Figure~\ref{fig:device-mgmt-cycle}, step c), the next stage is the \textit{deployment stage} (Figure~\ref{fig:device-mgmt-cycle}, step d), in which members of the community apply the firmware to a suitable target device and deploy the device at the desired location.
As soon as the device is deployed and is able to join the network, it is monitored by the platform (Section~\ref{sec:network-monitoring}).
In the \textit{monitoring stage} (Figure~\ref{fig:device-mgmt-cycle}, step e), the device is actively validated for correct operation in context of the whole network.
In case errors in configuration or other problems are detected, the device's maintainer is notified (Figure~\ref{fig:device-mgmt-cycle}, step f), so the device may be fixed and/or reconfigured, repeating the cycle.

The \nodewatcher{} platform aims to provide components for all stages of the described device management cycle, automating the repetitive tasks, freeing the human resources of the community and lowering the required entry technical skills level for active participation in the community network.  
Each part is designed to be easily extensible to networks with various topologies, routing protocols, operating systems and hardware devices.

\subsection{Platform-independent Configuration}
\label{sec:platform-independent-configuration}

\begin{figure}
\centering
\begin{lstlisting}[language=Python,frame=single,basicstyle=\ttfamily\footnotesize,breaklines=true]
# Module A
class InfoCfg(RegistryItem):
  name = CharField()

# Register schema item into the schema which makes
# it available to any other module.
registration.point('node.config').register_item(
  InfoCfg
)

# Module B
class ExtendedCfg(module_a.InfoCfg):
  device = ChoiceField('core.general#device')
  version = IntegerField()

registration.point('node.config').register_item(
  ExtendedCfg
)
\end{lstlisting}
\caption{A simplified example of a schema definition for a node's name and used device hardware, split over two modules to show extension capabilities.}
\label{fig:schema-node-general}
\end{figure}

Community networks are built using a wide range of devices, containing everything from off-the-shelf home routers to specialized devices used for backbone links and regular servers.
The \nodewatcher{} uses an extensible platform-independent schema to describe configuration for all these types of nodes, regardless of their hardware and/or operating system.
One of the motivations behind this choice is that platform-independent configuration enables replacement of devices without the need to do re-configuration even when a replacement device is of a different model or even manufacturer.
It is a frequent occurrence that deployed devices need to be replaced when they stop functioning properly due to various hardware failures.
Instead of taking down the hardware device, spending time on the roof or in the laboratory to fix it, and then taking it back to install it, with extended downtime during that period, it is better that the device is immediately replaced with a new one, but with exactly the same configuration as the previous one had.
The broken device can then be fixed without hurry and reused at some other location.
Because community networks are mostly operated by volunteers, dealing with urgent matters adds extra pressure on volunteers for their scarce time.
Additionally, replacement of a broken hardware device can be done easily by one volunteer, while fixing a device can be done at a later time by another volunteer.
Obtaining exactly the same device for a replacement is hard in community networks which utilize diverse hardware devices, so having a way to apply the same configuration to a new and different hardware device is needed.

\begin{figure}
  \centering
  \includegraphics[scale=0.47]{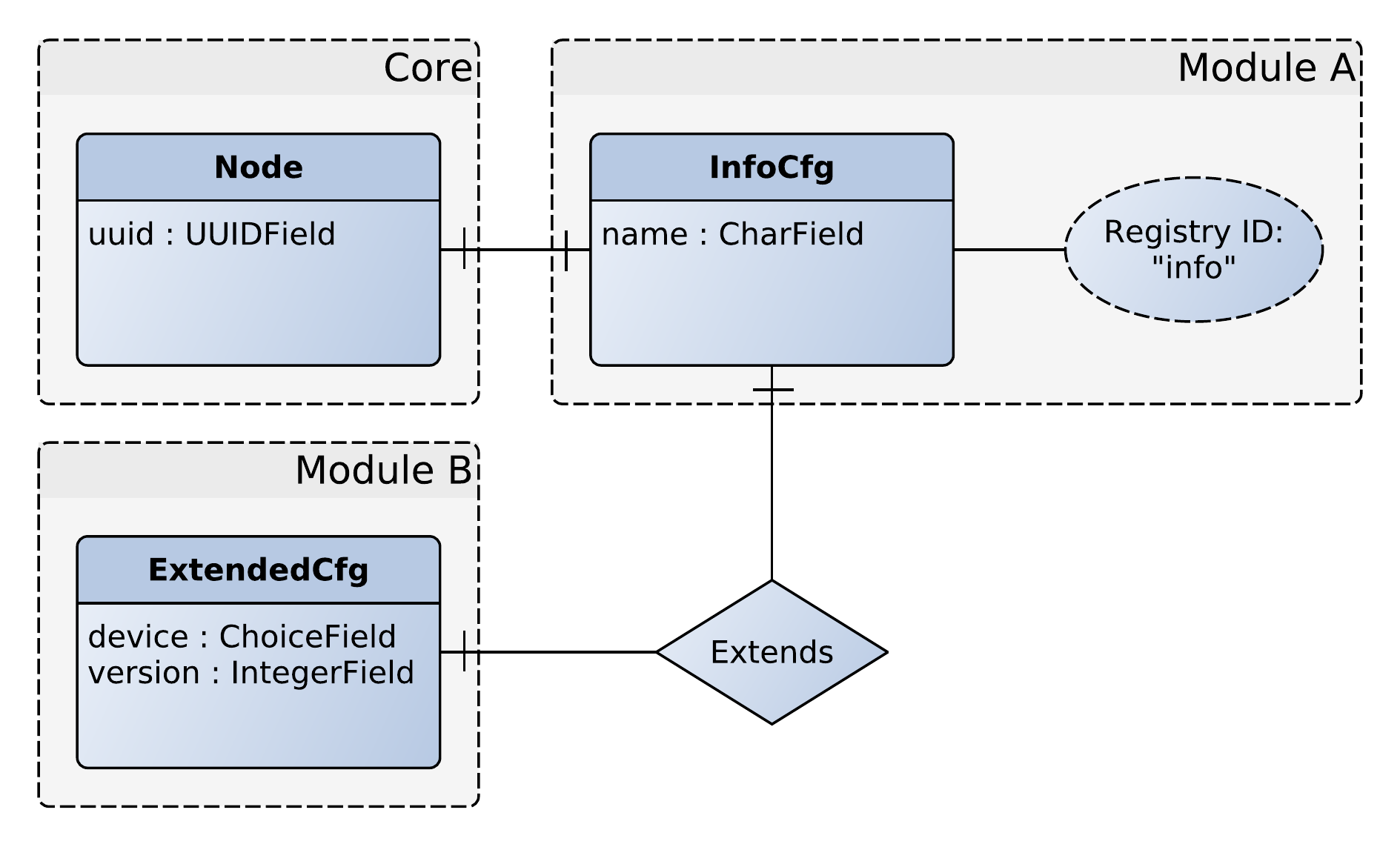}
  \caption{Schema item definition from Figure~\ref{fig:schema-node-general}, displayed using an entity relationship diagram.
  The \textit{core} module only contains a minimal \texttt{Node} while any attributes are provided by attached schema items in other modules.}
  \label{fig:registry-schema-example}
\end{figure}

However, while having a platform-independent configuration is a noble goal, some configuration properties depend on features which are inherently device-dependent (for example the number of Ethernet ports, available wireless radios, supported protocols, etc.).
In such cases the user editing the platform-independent configuration may create a configuration which will fail to work when applied to the target device.
This can further delay problem discovery until the \textit{deployment} stage when it is already too late and costly to fix problems, especially in wireless networks where nodes may be deployed in hard-to-reach locations.
This clearly shows the need to have instant validation and feedback (step c in Figure~\ref{fig:device-mgmt-cycle}) when updating platform-independent configuration.
Such validation must be based on the selected target device with all its hardware and software properties.
The \nodewatcher{} enables instant validation which is handled by the firmware generator component (see Section~\ref{sec:firmware-generator}).

In the introduction we have mentioned the problem with attempting to design an all-encompassing schema or a single node database application that would cover every possible deployment of community networks.
Communities will usually have some specifics regarding their operations~-- for example, because of different local regulations, hardware availability, or differences in philosophy.
Having a single unified schema can quickly become a limiting factor that prohibits straightforward adaptation of the system for the local community.

\nodewatcher{} avoids this problem by making the platform-independent schema itself completely extensible.
Individual modules may register schema items and the final schema is the union of all these items.
An example of a schema item definition is shown in Figure~\ref{fig:schema-node-general}, where several properties of the schema extension mechanism can be shown:
\begin{itemize}
    \item The schema items are class-based, which means they can be extended later on by other modules (in the example, \texttt{ExtendedCfg} from module B augments a simpler \texttt{InfoCfg} provided by module A to insert additional fields).
    \item Fields that represent enumerations (in the example \texttt{device} is such a field) do not hard-code the possible options, but only provide an \textit{extension point} where additional choices can later be registered by other modules.
    Each such extension point is attached to a unique name (e.g. \texttt{core.general\#device}) that may be referenced later when extending it.
    \item Schema items may also reference other items in order to build hierarchical configuration trees. This functionality may be used to define relationships, for example when a single wireless radio, defined by a schema item, may contain multiple virtual network interfaces with their own ESSID configurations, defined by other schema items (not shown in the example due to limited space).
\end{itemize}

The system that supports such schema item registrations is called the \textit{registry}.
It is essentially a lightweight extension of the standard object-relational mapper (ORM) concept~\cite{Bernstein_2007,ONeil_2008} as the mentioned schema items are actually database models.
In the platform, it is implemented using a popular Django web framework ORM \cite{django_2005}.
The problem that it aims to solve is the one of simplified model discovery.

In an extensible platform like \nodewatcher{}, modules may want to query on fields defined by other modules somewhere in the schema.
Reusing the previous example, there may be a schema item called \texttt{InfoCfg} in the base schema, enabling users to configure a \texttt{name} for a node (an entity relationship diagram corresponding to schema definition from Figure~\ref{fig:schema-node-general} is shown in Figure~\ref{fig:registry-schema-example}).
This base item does not provide any other fields besides the node name, leaving potential extensions to other modules.
Suppose that we would like to also add two more fields which would specify what device is in use on a specific node using an extensible choice field and some version information (in reality, version specification is more complex than a single integer, but we simplify this for our example).
We may do that in another module by defining \texttt{ExtendedCfg} as shown in Figure~\ref{fig:schema-node-general}.
Now, another module would like to perform a query, listing only nodes that use a device called \texttt{tp-wr741ndv4}.
As mentioned, these two classes actually represent ORM model definitions, so a traditional query traversing these two relations could be written using the SQL-like relational query notation introduced in~\cite{ONeil_2008}:
\begin{lstlisting}[language=sql,frame=single,basicstyle=\ttfamily\footnotesize,breaklines=true]
SELECT n FROM Node n
WHERE n.infocfg.extendedcfg.device = 'tp-wr741ndv4'
\end{lstlisting}

Here we make an assumption that there is a one-to-one relation between a \texttt{Node} and \texttt{InfoCfg}.
While one-to-many relations are also supported (so each node can have multiple instances of a model, for example multiple network interfaces), we limit ourselves to this simple case for ease of exposition.
Note how even in this very simple example we had to explicitly specify the hierarchical path that is spanning these two relations to get to the wanted field.
This makes such a method not very developer-friendly under the requirement of extensibility and even more complex schemas.
One of the features that the registry enables, is that we can simplify the same query as:
\begin{lstlisting}[language=sql,frame=single,basicstyle=\ttfamily\footnotesize,breaklines=true]
SELECT n FROM Node n
WHERE REGISTRY info.device = 'tp-wr741ndv4'
\end{lstlisting}

In this case, \texttt{info} is the \textit{registry identifier} that we attached to the base model using its meta \textit{Registry ID} attribute (see Figure~\ref{fig:registry-schema-example}) and represents itself or any of its subclasses at the same time.
In the background, the proper relations are automatically deduced and the query executed.
This abstracts away the subclass relationships, enabling easier refactoring and improving code readability.
Similar extensions are also provided for simplified fetching of fields deeply nested in the schema by deducing and performing the required table join operations.

In order to bootstrap module development, \nodewatcher{} already provides a minimal base schema for platform-independent node configuration.
To construct it, we have surveyed all the different solutions mentioned in Section~\ref{sec:related-work} and included a small amount of items common to most of the communities.
But even these base schema items are just module registrations which makes them removable in case some community needs to really change the baseline of how configuration is organized.
Using such an approach means that communities are not forced to adapt to a specific platform.
Instead, it empowers them to adapt the platform for their own use cases.
In case such modifications show themselves to also benefit other communities, they may also be more easily reused due to their modular nature.

\subsubsection{Extensible User Interface}
\label{sec:form-generation}

\begin{figure}[t]
  \centering
  \includegraphics[scale=0.5]{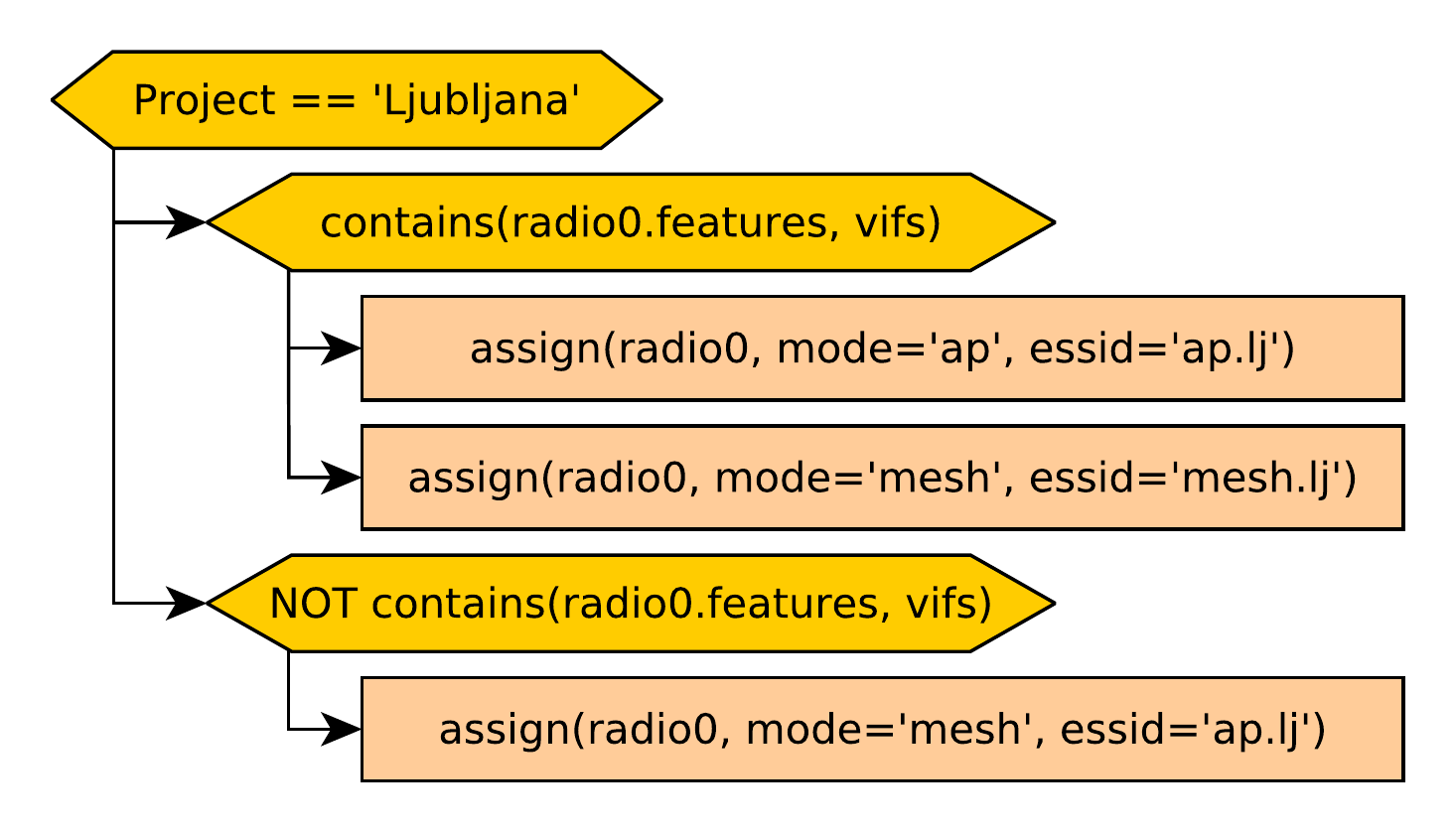}
  \caption{Example showing a compacted expression tree for specification of context-sensitive defaults for a wireless ESSID configuration. Hexagons are rule condition expressions while squares are action expressions.}
  \label{fig:defaults-rules-example}
\end{figure}

Having an extensible configuration schema is a good step towards reusable modules shared between community networks.
But, a schema is only useful for module developers that need a place to store various configuration values.
For platform users, the frontend (web interface) is even more important.

There are two issues regarding user interaction with the configuration schema:
(\textit{a}) there must be a way for the users to enter configuration values conforming to the specified schema;
(\textit{b}) since the schema may be complex, there needs to be a way for project maintainers to be able to specify context-sensitive configuration defaults.
Issue (\textit{a}) is addressed in \nodewatcher{} by the registry API's ability to automatically generate a user interface (forms) conforming to the schema.
Automatic form generation simplifies the module development process and reduces code duplication.
The automatically generated forms may be customized by the module developers where needed, but even defaults are immediately usable for simple schema items.
Addressing issue (\textit{b}) is especially important in order for community networks to be more accessible to people that do not have all the deep technical knowledge on how to configure devices.
Having the ability to define sensible defaults for such users is a feature that enables the community to grow by also including them.
A possible solution would be for network maintainers to provide pre-defined templates of configuration defaults.
The problem with static templates is that defaults may differ when applied to devices with different capabilities or a different project, in other words the defaults may be context-sensitive.
One would then be required to create static templates for all combinations, which quickly becomes unmanageable due to a combinatorial explosion in the number of required templates.

\nodewatcher{} takes a different approach, enabling specification of defaults in the form of simple declarative rules.
The example in Figure~\ref{fig:defaults-rules-example} shows context-sensitive default wireless ESSID configuration.
Rules in this example only apply to a specific project and configure two virtual interfaces (VIFs), one in mesh mode and the other in access point mode, in case the radio supports them.
In case the radio does not support configuring virtual interfaces, only one network may be set, and in this case a single mesh mode interface is configured.
The benefit of using a declarative approach for specifying rules instead of an imperative one is that the rules may only be evaluated when needed~-- in the above example, the inner rules will be evaluated only when the project changes and not when any other fields in the schema change.
This is an important detail as defaults should not overwrite configuration when changing an unrelated setting.
Evaluation of these rules is achieved by first generating an expression tree and then lazily evaluating only those sub-trees which contain expressions that match the current configuration value change.
In order to detect which rules have already been evaluated, forms generated by the registry contain internal state.

Automatic form generation combined with context-sensitive defaults also enables the possibility of generating different forms for the same schema, for example, depending on user settings or permissions.
Novice users may otherwise get confused by the sheer amount of things to configure, so simplified configuration forms may be shown to them while defaults are configured in the background.

In addition to configuration forms, \nodewatcher{} also includes interfaces, which allow developers to build upon modularity when they are reading, displaying and/or visualizing data for the user.
We extended the Django templating engine to allow modules to override or extend existing templates in a cascading way~\cite{Overextend_2013}.
By structuring templates appropriately, this enables developers to augment or change any server-side rendered content.
They can modify and define new menus, add new partials, or wrap or change existing ones.
In a similar way how we can automatically render forms based on the registry ORM, we can provide an automatically generated REST endpoints/APIs for all the data that is stored in these schemas.
As a consequence of all this, the community can quickly prototype new solutions over the existing base, or just use the exposed REST APIs in their own custom solutions or frameworks.

\begin{figure*}
  \centering
  \includegraphics[scale=0.5]{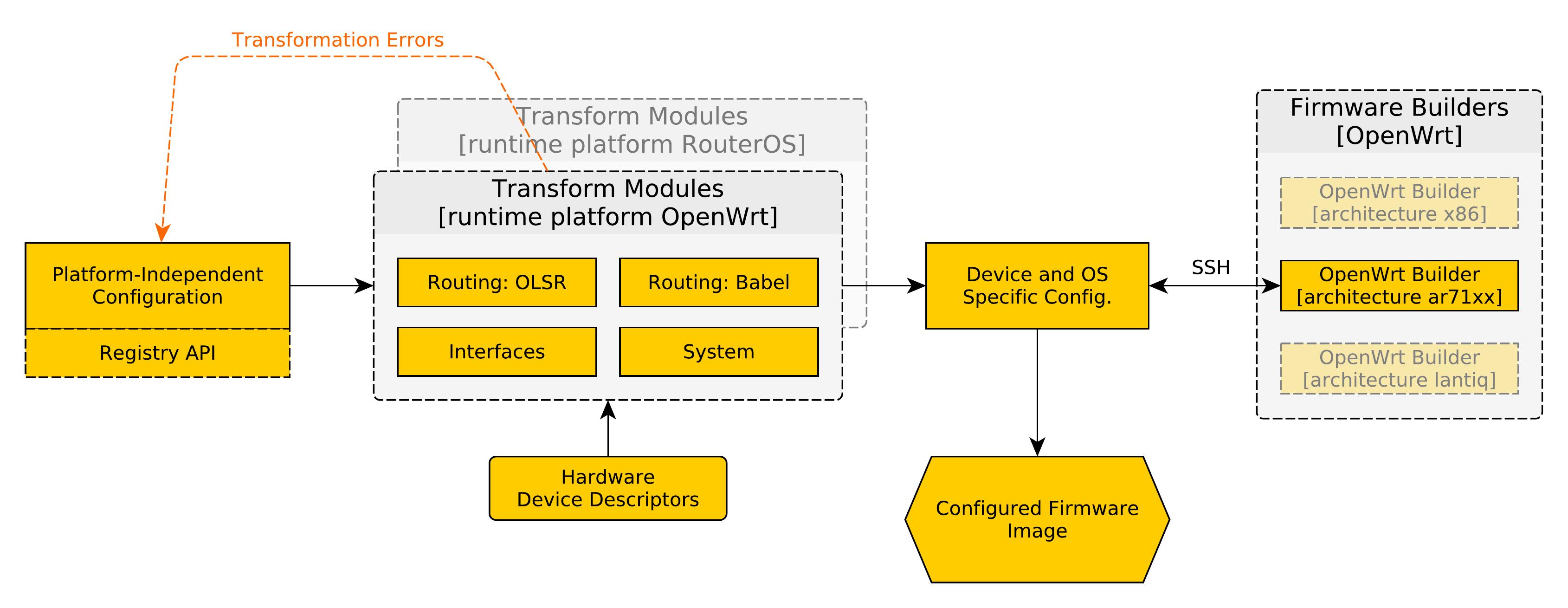}
  \caption{An overview of the firmware build system, coming from platform-independent configuration in the first stage to the fully configured firmware image that may be flashed directly onto the target device in the final stage.}
  \label{fig:firmware-build-system}
\end{figure*}

\subsubsection{Resource Allocation}
\label{sec:resource-allocation}

As in any network, there is also a need to perform IP address resource allocation in community networks.
This is especially the case in IPv4-based networks where it is hard to automatically generate node addresses without collisions due to the small available space.

In order to support that, \nodewatcher{} implements a hierarchical buddy allocation scheme~\cite{Peterson_1977} extended with support for hold down timers.
At the top level IP space is split into multiple pools from which other objects (for example nodes) may request specific allocations.
Hold down timers are necessary to avoid collisions with nodes that have been recently removed from the database.
When an allocation is freed, it is still marked as \textit{reserved} until the hold down timer expires.

This is necessary especially in community networks where there is limited coordination between volunteers.
Nodes may be removed from the database, but not yet permanently removed from the network and may actually reappear at a later time, causing routing conflicts.

\subsection{Firmware Generator}
\label{sec:firmware-generator}

Traditionally, in the generation stage of the device management cycle, devices are configured manually before they are deployed.
This is usually done either through a command-line interface via telnet or secure shell (SSH) or a web-based user interface running on the device, depending on the device's firmware.
In both cases, however, this is an error-prone process due to manual user input.
Mistakes can easily happen and sometimes they might propagate to the deployment stage where they are hard and costly to fix.
In community wireless networks, devices are sometimes deployed in hard-to-reach locations like rooftops or high towers and fixing certain problems requires physical access to the device.

\subsubsection{Transforming Configuration}

As described in Section~\ref{sec:platform-independent-configuration}, \nodewatcher{} can be used to store device configurations in a platform-independent way using the schema items exposed by the registry.
But this configuration cannot directly be used on devices.
Different operating systems like the open source OpenWrt~\cite{OpenWrt_2004} or the proprietary RouterOS~\cite{RouterOS_1995}, that are frequently used on devices in community networks, have completely different ways of being configured.
This is why an additional platform-dependent transformation step is needed.

However, such a step introduces some additional challenges that prevent a straightforward transformation of any configuration based on the platform-independent schema to a device-specific one.
This is due to the fact that there are differences between operating systems which may prevent certain configurations from being properly instantiated on one operating system even when those same configurations work without issues on another one.
Additionally even using the same operating system, devices have different capabilities due to differences in their hardware.
Configuration which was platform-independent and unaware of the target device in the first stage can produce problems while being applied to a specific device and operating system.
This conflict may result in some unpleasant, but realistic, scenarios:
\begin{enumerate}[label=\roman*)]
\item Devices have different default network switch layouts, VLAN tags and interface names.
As usually nodes use the WAN-designated port for the internet uplink and the LAN-designated port for routing to nodes in the same location, such a misconfiguration will cause connectivity issues.

\item Configuration of some wireless authentication mechanisms requires the installation of specialized packages on some operating systems.
Without them even an otherwise valid configuration will not work.

\item Different devices have different radio capabilities.
For example, some devices only support IEEE802.11a channels and if the configuration system is not aware of this, blind configuration transformation to the target platform results in a failure to bring up the wireless device.
\end{enumerate}

These scenarios show that supporting informed decisions of the configuration transformation process requires the use of a {\em device database}.
\nodewatcher{} takes a declarative approach to device descriptors which enumerate all the hardware and software properties of a given device.
A declarative descriptor is composed from multiple fields as follows:
\begin{itemize}
\item A unique model identifier in the form of \texttt{tp-wr741ndv1} which identifies a specific version of a device model in the database. The version is part of the identifier because there may be multiple hardware revisions of the same model, often with substantial differences.

\item A name used for representation of the device in user interfaces together with the name of the manufacturer and a URL to its website.

\item The hardware architecture (e.g. \texttt{ar71xx}) used by the CPU of the device.

\item A list of wireless radios embedded on the device.
Each radio contains a unique identifier, a name and a list of supported IEEE802.11 protocols and features.
It also contains a list of physical antenna connectors so that the configuration system may know to which port a specific antenna is attached in case there are multiple antenna ports available on the device.

\item A list of Ethernet switches connected to the CPU together with VLAN tag information and designation of the ports connected directly to the CPU.

\item A list of Ethernet interfaces and their connections to declared switches together with their VLAN tag definitions. 
Different devices have different switch and Ethernet port configurations and these two options abstract all of this into simpler identifiers like \texttt{lan0} and \texttt{wan0}.

\item A list of antennas that are included in the device package.
Each antenna contains radio propagation properties which characterize the antenna.
\end{itemize}

Device descriptors may be subclassed in order to simplify definitions of new devices with only slight variations.
This feature greatly improves the time to fully support new devices which is important in the quickly evolving community networks.

Using device descriptors and the platform-independent configuration provided in the first stage, \nodewatcher{} is able to generate device-specific configuration using a transformation step (see Figure~\ref{fig:firmware-build-system} for an overview of the whole process).
The transformation step is built from a pipeline of modules where each of them gets the platform-independent configuration as input and may produce modifications to the device-specific configuration as its output.
Such a modular transformation step ensures that the pipeline can be adapted to a wide range of transformations (for example, supporting various routing protocols, sensor inputs, network configurations etc.), so everything that a target device and operating system support may be used.
The transformation module pipeline may also raise errors when parts of the input configuration could not be properly transformed. These errors are immediately visible to the user who is entering configuration via the \nodewatcher{}'s web interface (as a part of step c in Figure~\ref{fig:device-mgmt-cycle}).
The validation system will not save a configuration which has outstanding errors, preventing invalid configurations from being used to deploy devices.

In order to further ease deployment the resulting device-specific configuration can be automatically packaged together with a firmware image which can then be flashed directly onto the target device.
Pre-generated firmware images further reduce the room for errors as no configuration needs to be transferred separately or entered manually.
After flashing devices boot directly into the configured and known state, ready to be used and deployed.
Besides reducing errors, packaging software (firmware) together with its configuration is also beneficial for ensuring that the configuration really is applicable to the used software versions as both can be tested together.
Otherwise, using a stale configuration on a newer operating system or newer versions of some packages may result in failing devices.

Bundling firmware together with node-specific configuration is a novel way of provisioning devices in community networks that is beneficial to existing and emerging networks alike.
It makes the whole device preparation process fast which allows volunteers to focus their time on deployment of the device on location.
Moreover, while experienced volunteers who often help maintaining multiple nodes gain the benefit of this streamlined node preparation process which minimizes time spent and possible errors, the important advantage is that the whole process becomes accessible also to the novice volunteers who are preparing their device for the first time.
They can register a new node, use provided defaults, select the hardware target, generate the image, flash it onto the device, reboot, and the device is ready to be used.

Moreover, the process of node preparation is repeatable and can be easily duplicated which helps debugging any potential issues.
If issues with a deployed node are observed, instead of having to retrieve the device and debug it, a new device with exactly the same configuration and firmware can be recreated in a laboratory environment.

\begin{figure*}
  \centering
  \includegraphics[scale=0.5]{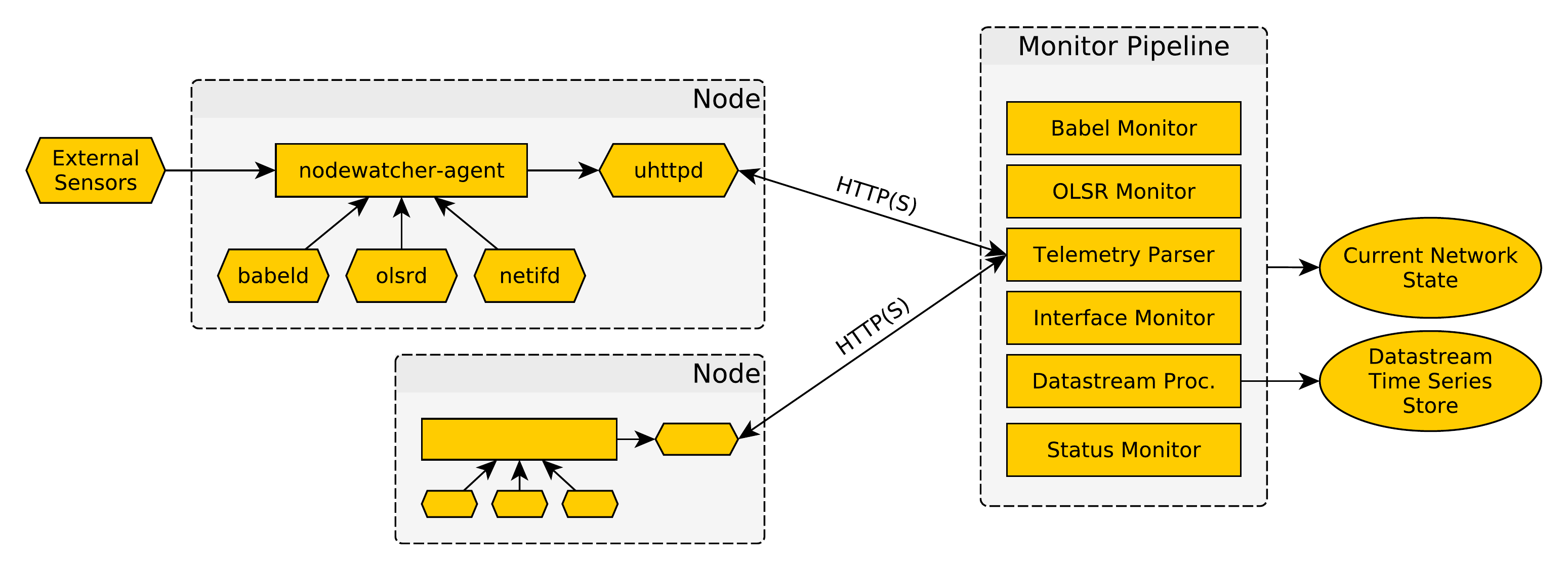}
  \caption{An overview of the \nodewatcher{} monitoring components. Telemetry data is collected on the devices by \texttt{nodewatcher-agent} modules and is then transported over HTTP(S) to the monitoring pipeline. This can be done either by pulling from the nodes or by the nodes pushing data to nodewatcher. One of the modules in the pipeline is the datastream processor which stores all the historical data as time series, supporting later interactive visualization through time.}
  \label{fig:monitoring-pipeline}
\end{figure*}

\subsubsection{Supporting Hardware Diversity}

Different community networks use various hardware devices. Additionally, new hardware is being developed all the time.
A network management platform is therefore only really useful if it enables easy inclusion of new devices and even new operating systems.
To enable this, \nodewatcher{} splits hardware support into different components, which may be provided by independent modules:

\begin{description}
    \item[Runtime platform.] The runtime platform is dependent upon the operating system that runs on the target device.
    OpenWrt and RouterOS are examples of runtime platforms.
    But there is no hardcoded concept of how a platform should behave.
    What defines the runtime platform are the transformation modules, which contain the logic of converting the platform-independent configuration into something that can be understood by the target device.

    \item[Firmware builders.] Separate from the runtime platform the system includes one or more firmware builders.
    Each contains a set of tools, which are able to generate firmware images that may be copied directly onto the target device (also called a {\em toolchain}).
    In case the devices are using proprietary operating systems, these may not even exist.
    Decoupling the builders from the runtime platform means that proprietary runtime platforms can also be supported.
    In such cases, configuration will still need to be applied manually.

    \item[Hardware device descriptors.] These are the device descriptors that we have already defined.
    They provide to the transformation modules the knowledge required to correctly adapt the platform-dependent configuration to the target device.
    Since a device may in theory support different runtime platforms (e.g. some Mikrotik devices may use either OpenWrt or RouterOS), the same device descriptor can be reused by multiple runtime platforms.
    In this case, the platform-specific properties are specified for each platform, while the common ones are specified only once for each device.
\end{description}

As seen in Figure~\ref{fig:firmware-build-system}, firmware builders are kept separate from the runtime platform, so that proprietary systems may be supported.
The link between the runtime platform and the firmware builders is the platform-specific configuration.
This configuration is the output of the transformation modules defined for the target runtime platform.

The tools used to build firmware images for embedded devices can be complex and may vary wildly between the runtime platforms.
This is why the firmware build system in \nodewatcher{} has been designed in such a way that it can be used by completely different sets of tools in a modular fashion.
The build system is structured into multiple Docker containers~\cite{Docker_2013}. Docker containers are a lightweight wrapper around the Linux namespacing API and filesystem layers with a goal to enable an interface for packaging applications in a reusable and extensible way. Namespaces provide container isolation (the containers still share the host kernel), so that adjacent containers running on the same host are unable to see or influence each other's processes, network configuration, etc. Each container can be thought of as a very lightweight virtual instance, but without the overhead of running a full virtual machine with its own kernel.
\nodewatcher{} uses the Docker container features in order to generate and run firmware image builders for multiple runtime platforms.

After the platform-specific configuration has been generated by the transformation step of the given runtime platform (see the node \textit{Device and OS Specific Configuration} in Figure~\ref{fig:firmware-build-system}), it is directed to the suitable builder that is selected based on the hardware architecture specified in the device descriptor.
Once the generated firmware images are prepared, they are made available to the user via the \nodewatcher{} frontend. This decoupling of platform-specific configuration and firmware builders enables the builder containers to be distributed and deployed on a cluster of machines to better handle the load, resource availability and utilization.
The described system is also extensible~-- adding support for new architectures simply requires a new builder container to be prepared, while adding support for new runtime platforms also requires an extension of the configuration transformation modules.

But the main advantage of using separate containers for individual firmware builders is that it enables simple reuse and sharing of ready-made builders between community networks, which is one of principles of community networks.
Compiling and preparing a range of firmware builder images can be a lengthy and resource-intensive process, requiring manual configuration and testing.
Containerized organization enables communities to simply download pre-built builder images and use them in their \nodewatcher{} installations or \nodewatcher{}-compatible systems without needing to compile anything.
This further simplifies the bootstrapping of new community networks which can simply reuse firmware builders from existing community networks.

\subsection{Network Monitoring}
\label{sec:network-monitoring}

After the firmware images are prepared and devices are deployed in the field, we now enter the monitoring stage of the device management cycle.
In this stage we constantly monitor the devices for status, performance and compliance.
In the same way as the configuration transformation step performs static validation of device configuration before it is deployed, the role of the monitoring component is to perform dynamic validation of device configuration after the device is running.
While validation is common to both, the scope of the latter is much bigger~-- when deployed in a large network, the functioning of one node may also affect other nodes in its vicinity or sometimes even in completely different parts of the network (for example when considering network announce conflicts in routing protocols).
Besides performing configuration compliance validation, monitoring may also be used to collect various sensor data through time.
This is useful for diagnostics under changing network conditions and can also be used to collect sensor data coming from external sources like temperature, humidity and lightning strike detection sensors.

One may ask why should monitoring be integrated into the provisioning platform?
It is true that existing network monitoring tools could easily be used instead, but an integrated solution enables the monitoring modules to easily perform validation of current device state and behaviour against the static platform-independent configuration that is stored in the provisioning database.
This enables the system to quickly detect configuration errors (for example after someone manually edits configuration on a device) or failure modes (loss of a redundant VPN link only when such a redundant link has been previously configured).
Such capabilities could be replicated using existing systems, but this would require either manual duplication of configuration (an error-prone process) or specific import scripts for the target monitoring system (which is usually not very portable among different communities).
Thus, an integrated monitoring component is key in ensuring ease of deployment and transfer of good practices between community networks in the form of modules implementing the validation procedures.

\begin{figure}[t]
\centering
\begin{lstlisting}[frame=single,basicstyle=\ttfamily\footnotesize,breaklines=true]
{
  "core.general": {
    "_meta": { "version": 4 },
    
    "uuid": "64840ad9-aac1-4494-b4d1-9de5d8cbedd9",
    "hostname": "test-4",
  },
  "core.resources": {
    "_meta": { "version": 2 },
    
    "memory": {
      "total": 32768,
      "free": 24611
    }
  }
}
\end{lstlisting}
\caption{Example part of the JSON schema compiled by the \nodewatcher{} monitoring agent, showing sample output for two monitoring modules.}
\label{fig:monitoring-json-schema}
\end{figure}

In traditional computer networks, especially of such scale, provisioning of network devices would be done automatically and centrally.
There would be little reason to assume misconfiguration of nodes.
Most issues with operation would be because of software or hardware bugs, hardware failure, or human error from operations control.
In any case, the misconfiguration would be easy to address: simply provision the node again.
On the other hand, most community networks see decentralized nature of their networks as an important aspect of their networks and do not want a centralized and automatic control of nodes.
Nodes are often maintained by individuals who might use tools like \nodewatcher{} to streamline the process of maintenance, but they still want to retain control of and access to the device.
This can lead to potential issues, from having devices with old and potentially obsolete firmware versions in the network, to simply misconfigured nodes.
Such nature of community networks has to be taken into account and \nodewatcher{} helps detect any issues and guide maintainers towards resolving them.
They can use the provided firmware image, or can be guided through detected issues and suggested steps to resolve them manually.
For example, in \wlanslovenija{} network we had cases where maintainers connected devices with vanilla OpenWrt installed to the network.
\nodewatcher{} then detected invalid or missing configuration, provided instructions to resolve them, and maintainers manually vetted and followed them one by one until the node was brought into the compliance with the rest of the network.
Such approach is of course time-consuming, but it is an important part of community networks spirit to be able to retain complete control over the node and all aspects of its operation and configuration.

\subsubsection{Obtaining Telemetry Data}

An overview of the monitoring system is given in Figure~\ref{fig:monitoring-pipeline} which shows the data flowing from sources on the devices towards the time series data store and current network state as the sinks.
Data collection starts on a node and is implemented by the \texttt{nodewatcher-agent} process.
It is a small C application with a minimal core that is able to periodically request the loaded modules to provide their state updates which are then compiled into the current node status and exported in a JSON form.
There are then two ways for the agent to transfer the data to the nodewatcher backend:
\begin{itemize}
    \item \textbf{Pull.} The JSON data may be served over HTTP(S) and the nodewatcher monitoring backend will periodically request new data from the nodes.

    \item \textbf{Push.} The agent on the node will periodically push its monitoring data to the backend using HTTP(S) POST requests. This requires that the push URL and interval be configured on the node.
\end{itemize}

This behaviour of the agent may be configured per-node.
Some nodes may push data while data is pulled from others.
Supporting both modes of operation is beneficial for situations where the \nodewatcher{} backend installation does not directly see every node in the mesh network, but is instead located somewhere in the public Internet, without VPN access to the network itself.
With push support, nodes may provide telemetry data even in this case, by pushing data to a public URL.
As we have already mentioned, the agent is composed from multiple modules.
Each module is a shared library which is loaded when the agent process starts.
Having modules as shared libraries enables simple extension of the monitoring agent by third-party packages.
Modules may independently fetch data from external sources providing state like current resource usage reported by the Linux kernel, status of various interfaces, wireless configuration and site survey, connected clients, external sensor input, topology information obtained from routing daemons etc.

As can be seen from its description, the monitoring agent follows a similar modular design as other parts of \nodewatcher{}.
An important feature of a monitoring agent that runs on remote devices in a community network is the ability for modules to independently evolve their schema.
In order to add features to existing modules there needs to be a way to version the state schema which is reported back to the monitoring pipeline.
This is especially the case in community networks where there are many different devices with different firmware versions and also with different versions of the monitoring modules.
A single version for all modules is not enough as modules may be developed by different developers, possibly from different community networks and independent schema evolution is required.

Node state compiled by the agent is a structured JSON document where the top level contains one dictionary element for each module with the element's key being the module identifier.
A partial example of such a state is shown in Figure~\ref{fig:monitoring-json-schema}.
Each module element may provide whatever elements it wants to report for the current state.
The agent will automatically create a special \texttt{\_meta} element containing the module metadata~-- currently a module version number.
By inspecting the metadata, the processing pipeline is able to handle multiple versions of the schema for different modules.

The implemented agent that uses JSON over HTTP(S) connections is just one of the possible monitoring data source implementations.
The architecture enables other data collection protocols to be used side-by-side.
One possible such protocol, that many existing device operating systems already support, is SNMP.
While our custom protocol enables easier schema evolution through per-module versions, SNMP may be used in cases where custom monitoring agents cannot be installed on target devices.
This co-existence of data sources is enabled by the modular design of the monitoring backend.

\begin{algorithm}[t]
\begin{algorithmic}
\Procedure{MonitoringRun}{$P$}
  \State $W \gets \emptyset$\Comment{Initialize the working set.}
  \State $C \gets \emptyset$\Comment{Initialize the context.}
  \For{$p \in P$}\Comment{Iterate over the pipeline.}
    \If{$p \in P_n$}\Comment{Network processor.}
      \State $\langle W, C \rangle \gets p.\mathrm{process}(W, C)$
    \ElsIf{$p \in P_m$}\Comment{Node processor.}
      \For{$n \in W$}
        \State $C \gets p.\mathrm{process}(n, C)$
      \EndFor
    \EndIf
  \EndFor
\EndProcedure
\end{algorithmic}
\caption{A single monitoring run.}
\label{alg:monitoring-pipeline}
\end{algorithm}

\begin{figure}
  \centering
  \includegraphics[scale=0.4]{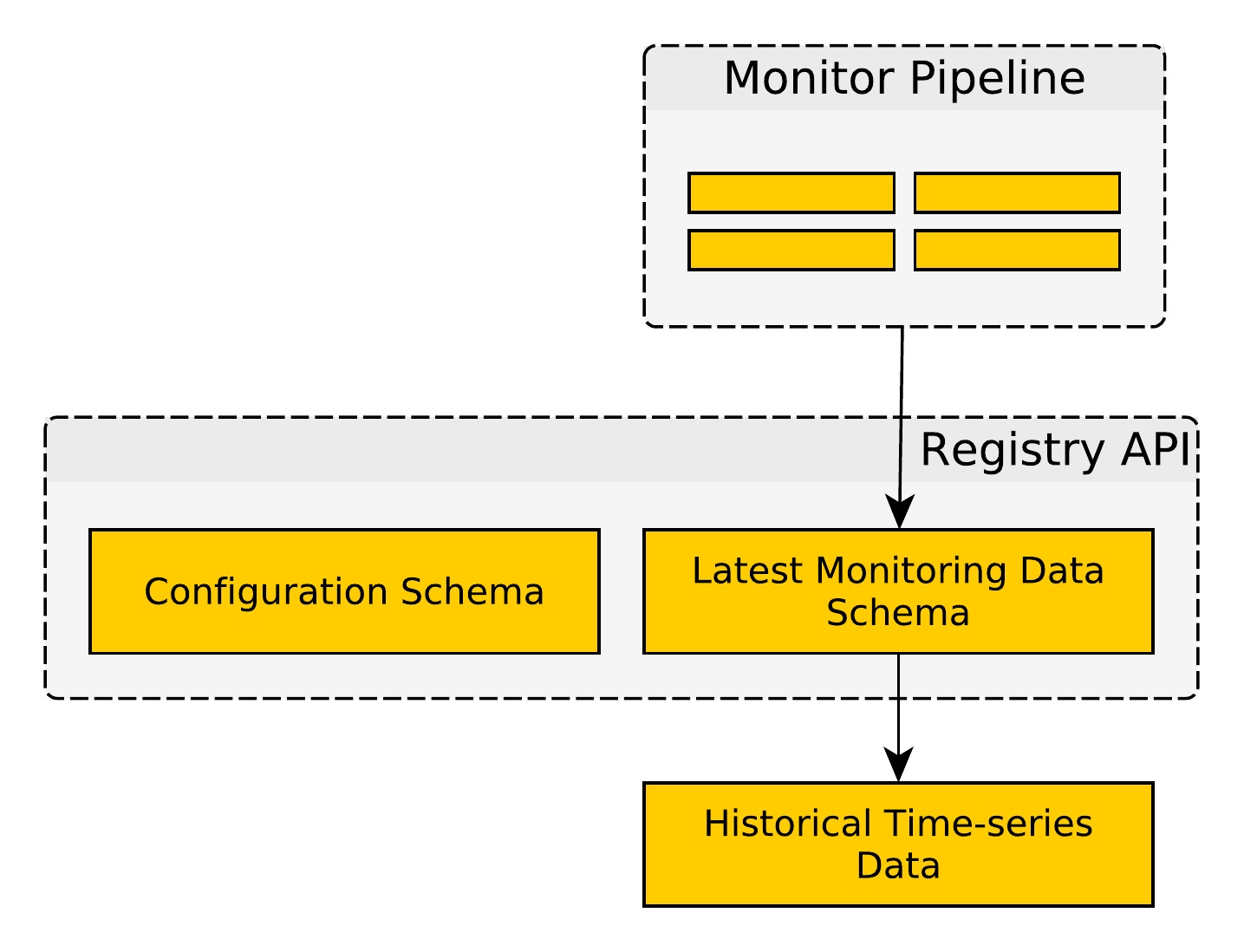}
  \caption{Overview of relationship between the configuration and monitoring data schemas and the flow of monitoring data from the pipeline to the historical time-series data storage.}
  \label{fig:storage-relationships}
\end{figure}

\subsubsection{Monitoring Pipelines}
On the backend, the processing of all operations related to monitoring is handled by the monitoring pipelines.
Conforming to the modular philosophy, the pipeline consists of processors which are implemented by modules.
Throughout the execution of the pipeline two pieces of state are maintained: a working set of node instances and a context.
The context can contain arbitrary structured data which is communicated between the different processors.
Working set of nodes represents the instances that next processors will operate on.
When execution of the pipeline begins, the working set is empty.
There are two basic types of processors:
\begin{description}
\item[Network processors $P_n$] are executed once with all nodes in the working set and context as arguments. They may modify both the working set and the context to change the flow of downstream processors.

\item[Node processors $P_m$] are executed for each node in the working set with context as an argument. They may modify the context but not the working set.
\end{description}

Algorithm~\ref{alg:monitoring-pipeline} shows a simplified version of the processing run execution.
As can be seen from the algorithm, the pipeline is completely generic and its content depends entirely on the processor implementations which are provided by modules.
In order to increase performance, the pipeline implementation actually performs some optimizations that cannot be observed in the above pseudocode.
Network processors must be executed sequentially as each may modify the working set which is part of the state for downstream processors.
But there is no reason why execution of node processors cannot be done in parallel for each node.
Additionally, if there are multiple consecutive node processors which will run on the same working set (note that only network processors may change the working set), they can all be executed inside the same thread, one after the other.
This greatly reduces the amount of context transfer between processes and speeds up the monitoring run execution.

The modular nature of the monitoring pipeline enables different communities to completely adapt it to their own network.
Since the base framework is common to all, modules from different community networks may interoperate inside the same pipeline, passing data through the context in a loosely coupled manner and increasing code reuse possibilities.

\subsubsection{Storing Monitored Data}

\begin{figure}
  \centering
  \includegraphics[scale=0.4]{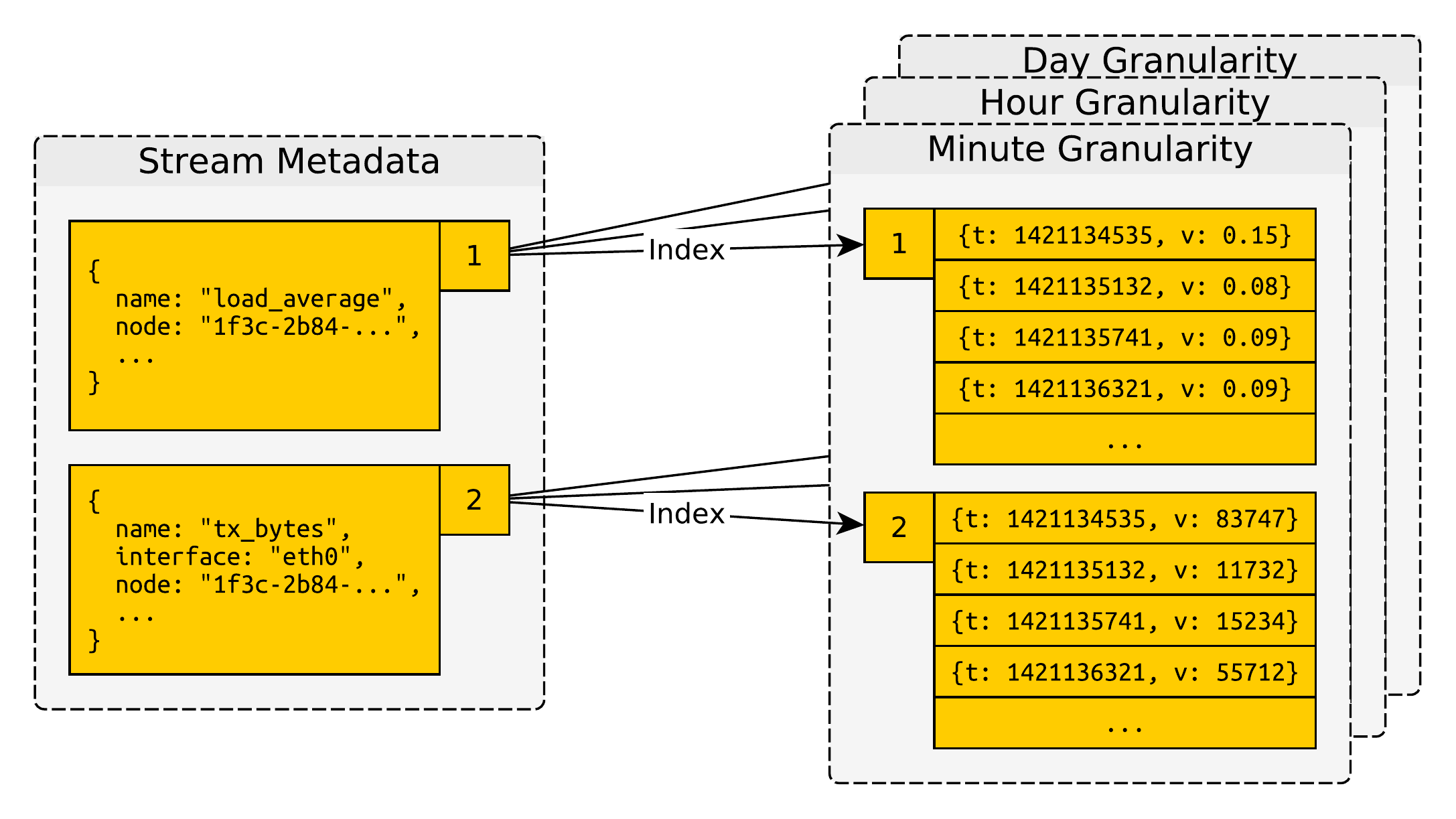}
  \caption{Datastream stream storage organization. Tag based metadata store compact indices into the stream data, downsampled at different granularities.}
  \label{fig:datastream-storage}
\end{figure}

\begin{figure}
  \centering
  \includegraphics[scale=0.4]{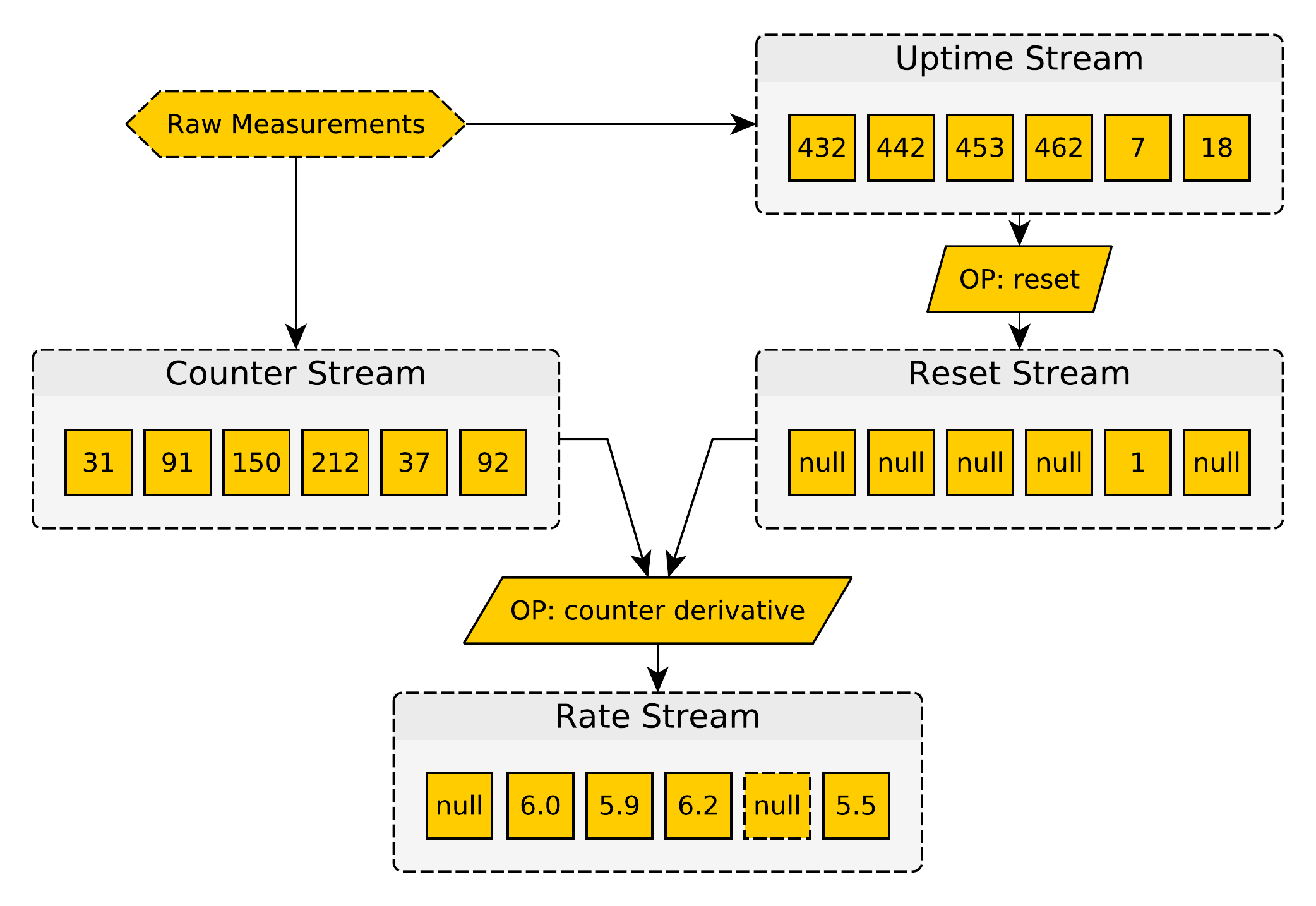}
  \caption{Correct rate computation example using the counter derivative operator, automatically derived from two streams.}
  \label{fig:datastream-counter-reset}
\end{figure}

In Section~\ref{sec:platform-independent-configuration} we described how \nodewatcher{} uses the registry to store nodes' platform-independent configuration, described by the configuration schema.
For storing monitoring data we use two different approaches that serve slightly different goals:
\begin{itemize}
\item Latest data obtained by the monitoring pipeline, describing the current network state, is stored in a similar form as configuration for each node.
In addition to a configuration schema, also a monitoring schema is defined and data is stored into a relational database in the form of schema items.
Reusing the same functionality, this automatically gives us the same schema item extensibility.
Data stored in this way may be queried by value and is thus used for performing configuration validation against observed data.

\item In addition to the above, we also want to keep historical data of how the network operated through time.
The storage and query requirements for historical data are completely different as these data are represented as time-series, indexed by time, not value.
Being a modular platform, the historical data storage is implemented as a module which may be added or removed as needed by the community.
\end{itemize}

An overview of the relationships between these storage components is shown in Figure~\ref{fig:storage-relationships}.
Latest data are sampled on every monitoring run and stored into the time-series data store.
The monitoring pipeline may generate large amounts of time-series data during its operation (for example \wlanslovenija{} has accumulated over 200 GiB of data in the last five years, storing everything from network diagnostics to external sensor data).
As shown in Figure~\ref{fig:monitoring-pipeline}, one of the processors in the monitoring pipeline can be a \textit{datastream} processor, storing time-series data.
Datastream is a system we developed, enabling storage, processing and retrieval of time-series data.
In contrast to round-robin databases~\cite{Oetiker_1999}, where the database size is fixed in advance and old data is simply discarded, we are taking a different approach, storing all the data that we can for possible later analysis, enabling future research on improving automatic network diagnostics.

\begin{figure*}
  \centering
  \includegraphics[scale=0.45]{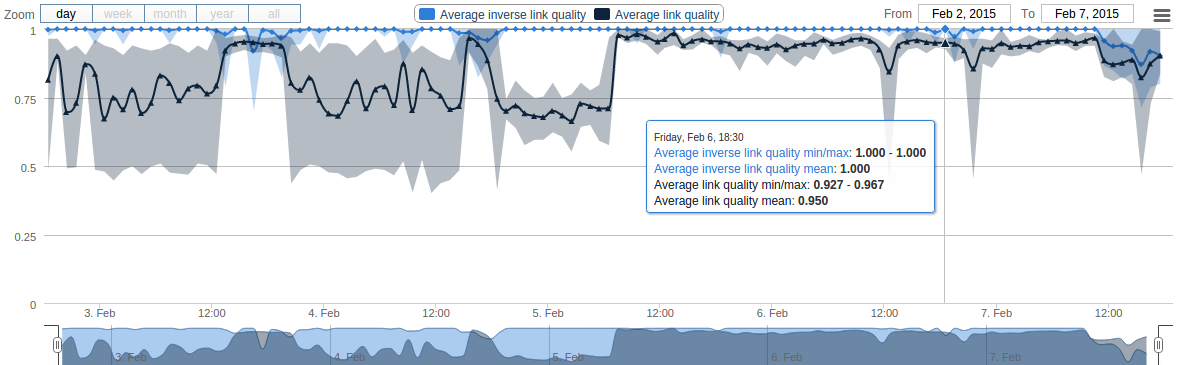}
  \caption{Interactive visualization frontend module in \nodewatcher{} for visualizing collected monitoring data from datastream storage allows various combinations of metrics to be visualized, enabling introspection into the functioning of the network.
  The visualization shows a plot of link quality, averaged over all links to neighbouring nodes, for a node in the mesh network as reported by the routing protocol over the course of five days.
  It displays the mean value together with minimum and maximum in the reported time interval which are all easily available due to downsampling.}
  \label{fig:interactive-visualization}
\end{figure*}

The database is built around a concept of append-only streams, each stream being an independent time series.
In order to organize the streams, each stream may be tagged using arbitrary key-value pairs (see Figure~\ref{fig:datastream-storage}), which are indexed and can be used for fast lookup of streams matching specific tags.
For stream storage to be efficient, datums are stored in separate collections, one for each time granularity, and connected with their metadata entries using compact indices.
The default storage backend used by datastream is TokuMX~\cite{TokuMX_2007}, a scalable document database based on MongoDB~\cite{MongoDB_2007}, which implements data compression and uses the fractal tree index~\cite{Brodal_2003,Bender_2007} data structure for indices, both of which greatly improve performance.
The design of datastream is modular, supporting implementations of alternative storage backends while continuing to expose the same API.
As other \nodewatcher{} components, it is available as an open source project~\cite{Datastream_2012}.

Each stream has some special tags which define its base operation.
Streams are typed, meaning that they may be used for storage of different data point types.
Supported types currently include numeric values (integers, floats, arbitrary precision numbers) and graph values (nodes and edges for storing how topologies evolve over time).
Streams also define the highest granularity at which data points will be inserted.
This setting is an optimization which allows the datastream storage backend to reduce the number of granularities when downsampling data points.
Datapoints are inserted only at the highest granularity and are then automatically downsampled in the background, making them ready for efficient querying and visualization.

Downsampling is a crucial component for ensuring that the system maintains good performance as the number of data points grows.
The motivation for downsampling comes from interactive time-series data visualizations which enable one to see the data at various zoom levels and for different time intervals.
For example, there is usually no need to display data points for every minute when looking at data in the interval of several months.
Even fetching this many data points from the database and sending it to the user's web browser may be overwhelming.
Downsampling aggregates higher-frequency data into buckets covering larger time intervals.
By design, any downsampling will cause loss of certain information in lower granularities.
In order to reduce the effect of information loss, datastream supports multiple aggregation functions that determine what data is preserved in downsampled versions of data points.
Aggregation functions include the point count, sum, sum of squares, mean, minimum, maximum and standard deviation, all computed over the data points allocated to the same time bucket.
Storing different statistical moments enables a better understanding of aggregated data and also enables improved visualizations where not only the average is shown (see Figure~\ref{fig:interactive-visualization}).

Streams may be automatically computed from other streams using different operators.
The most prominent use of this feature in monitoring is to support correct rate computations under the possibility of counter wraps and resets which could otherwise cause apparent rate spikes when the counter is reset due to a reboot but the system incorrectly classifies it as a counter wrap.
To illustrate why this can be a problem, imagine a simple 8-bit unsigned integer counter (its size is known to the monitoring platform), sampled every 10 seconds (see \textit{Counter Stream} in Figure~\ref{fig:datastream-counter-reset}).
Focusing on the instance where counter value decreases from $212$ to $37$ this can be treated either as a counter wrap (as its maximum value is $255$) or as a counter reset due to device rebooting.
Without additional information, such events must be classified uniformly: if they are all classified as wraps, rates may incorrectly spike (in our example the rate would be computed as $(255 - 212 + 37) / 10s = 8.0s^{-1}$); if they are always classified as resets, data points may be lost.

Using datastream, a \textit{counter derivative} operator accepts two streams, one containing raw counter data (for example the number of bytes transferred on a network interface) and one containing discrete events at which a device was rebooted (for example derived from its uptime by the \textit{reset operator}).
Using both pieces of information, the rate is computed as a derivative, but only when there was no reset event during the last time interval.
In case of a reset, a \texttt{null} value is inserted instead.
However, if there was no reset and the counter value decreased, the event is classified as a counter wrap and the rate is computed using the known maximum value of the counter.
Computing streams from other streams may be chained as shown in the example in Figure~\ref{fig:datastream-counter-reset}.
Device uptime, counting seconds from the device's boot time, is received as raw measurements which the \textit{reset} operator then transforms into a stream suitable for use in the counter derivative operator.
Such chaining naturally describes data transformations which happen while the data is streamed.
Additional operators include computing sums of multiple streams which is useful to compile aggregate traffic rates.

The datastream module also provides a REST API as a way to access current and historical time-series data with all the benefits of datastream data storage: quick access to datapoints at various granularities, diverse data types (numeric and graphs), easy navigation and searching for streams using custom tags which can also serve to guide visualization.
Using all these ways and general \nodewatcher{} modularity allows various ways of providing user interfaces to the data.
From server-side rendered templates, to dynamic JavaScript based visualizations like the one seen in Figure~\ref{fig:interactive-visualization}.
Because any data is provided through REST APIs as well it can be made easily exportable and shared with others.

\section{Evaluation and Discussion}
\label{sec:evaluation}

\begin{figure}
  \centering
  \includegraphics[scale=0.45]{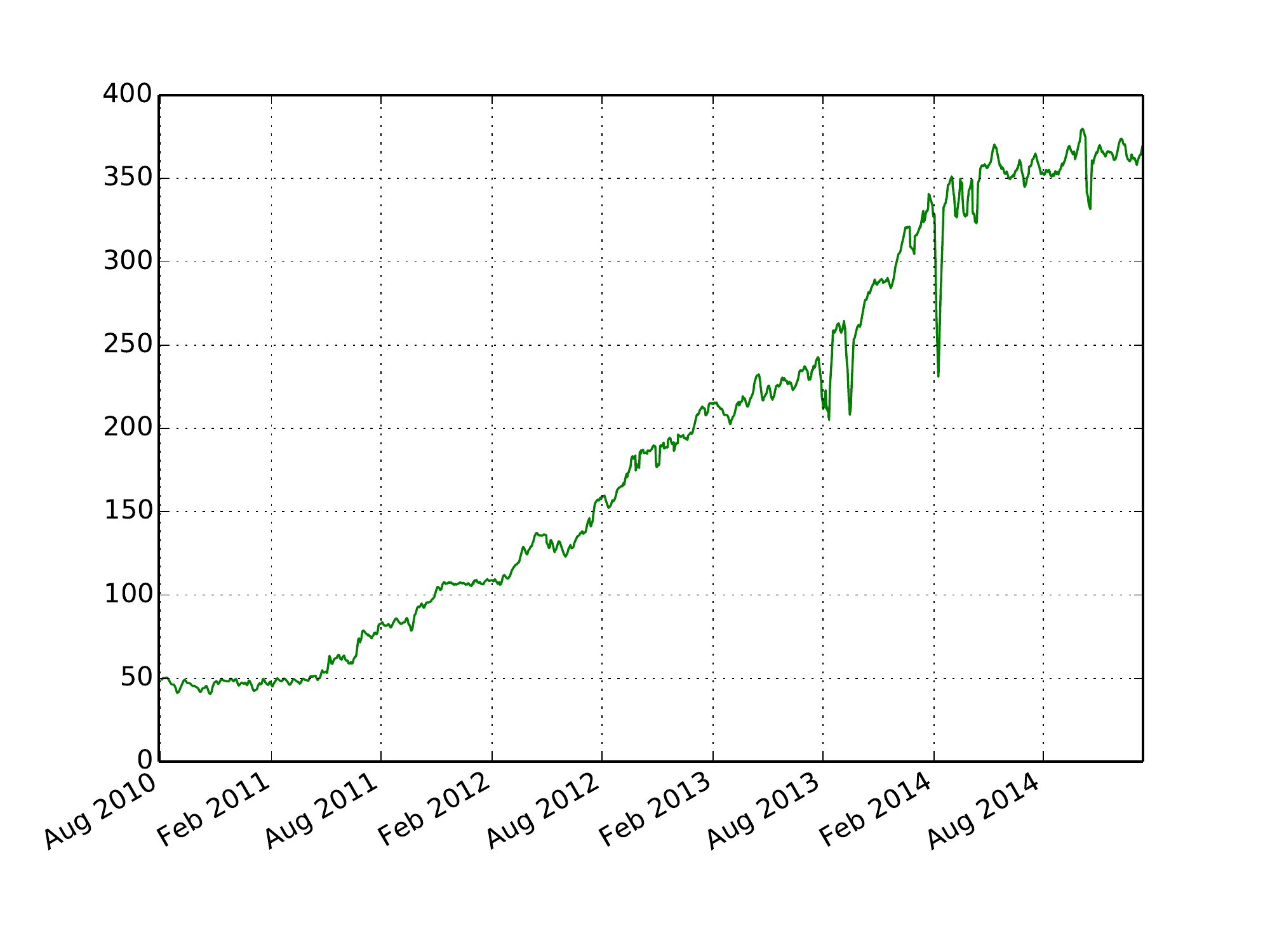}
  \caption{Number of online nodes in \wlanslovenija{} community wireless network as reported by the monitoring system for the past few years.}
  \label{fig:wlansi-nodes-up}
\end{figure}

In this section we evaluate \nodewatcher{} in the context of \wlanslovenija{} wireless community network.
\wlanslovenija{} has been an active network since the year 2009.
It is a medium-sized network with around $400$ online nodes at the moment (the growth of the number of online nodes since 2010 may be seen in Figure~\ref{fig:wlansi-nodes-up}).

Primary management of the network is currently still based on the older version v2 of \nodewatcher{}, while the new version presented in this paper is being run in parallel, since the beginning of 2015, in order to enable a smooth transition.
The old version is designed similarly to other existing community network platforms that we surveyed in Section~\ref{sec:related-work} and shares many of the same problems.
Due to this state we have a unique opportunity to qualitatively compare the workings of both solutions and show how the new version of the platform substantially improves network management by addressing the exposed problems.

\subsection{Device Support}

It is much easier to support new devices and keep up with the pace of requirements from other communities in the new platform.
Currently, our platform is also being used by the neighbouring community network in Croatia where they use some unique devices due to their cheap local availability.
In old version of the platform, every new device required substantial effort to support it properly as \nodewatcher{} required substantial changes to the provisioning system code.
Additionally, \nodewatcher{} did not have advanced validation capabilities, resulting several times in erroneous configurations which required re-flashing to fix.
In contrast, the new platform enabled us to quickly support the new device, simply by writing a new device descriptor for it (see Section~\ref{sec:firmware-generator}).

\subsection{Monitoring}

The new modular monitoring pipeline brings multiple improvements over the old monolithic version.
Measurements are performed faster and are easily parallelized over multiple cores due to node/network processor separation, as described in Section~\ref{sec:network-monitoring}.
A side effect of enabling parallel execution of processors is also support for performing monitoring runs at different intervals.
For example, topology information may be updated more quickly as it is readily and locally available from the OLSR routing protocol and does not require polling of all the nodes.
On the other side, device telemetry measurements require data requests over a wide network which may consume more time.
Even slower are the measurements of node reachability and packet loss using ICMP ECHO requests under varying packet sizes to check for MTU issues.
Previously, the slowest measurement affected the execution time of the whole monitoring run.
The new modular design enables such runs to be isolated so some may execute entirely in parallel and with higher frequency than other, slower, measurements.

Due to our new time-series data storage system, we are able to store a complete set of monitoring data, forever.
Not being limited by the fixed size of round-robin databases enables later analysis of network events over long periods of time.
Such analysis may enable new insights into how certain network problems are correlated~\cite{Steinder_2004}.
As we store both, configuration and latest monitoring data, we are able to perform comparisons and validate that the monitored operation matches exactly with what is configured for specific nodes.

\subsection{Interactivity}

The old version used a round-robin database for time-series data storage together with its visualization module which generated static images.
Static visualizations are limiting when it comes to having a good overview over multiple measurements.
Combining measurements over arbitrary time spans at arbitrary resolution greatly enhances the ability to diagnose problems.
Additionally, generating a large set of static images is very resource-intensive for the server.
Interactive visualizations, like the one seen in Figure~\ref{fig:interactive-visualization}, transfer only data points to the client machine and the web browser then performs all the rendering.
This naturally distributes the required computations and reduces strain on the central server.

\subsection{Modularity and Interoperability}

The Trac Project~\cite{Trac_2003} is an example of a modular system for managing open source projects.
Very early on in its development, a plugin system has been developed and hundreds of community-made plugins have been made since then, many available through their Trac Hacks~\cite{TracHacks_2004} repository.
We were inspired by it and decided to do something similar for managing community networks.
We built \nodewatcher{} using the Django framework~\cite{django_2005} and not a custom framework to leverage standard Django packages and modularity.
We can reuse package repositories and tools that the Django ecosystem provides.
For example, the Django Packages~\cite{DjangoPackages_2010} repository with more than 2600 packages available.

In 2014 interoperability efforts between community networks have been revived~\cite{interop_2010}.
Some very active communities started discussing and comparing data schemas used for configuration and monitoring.
During this ongoing process we have been re-evaluating schemas used by other communities and the resulting compromises stemming from the discussions.
Again and again we have been reaffirmed that we achieved our design goal because we are discovering that \nodewatcher{} can support all of the proposed schemas with little effort and are actually able to quickly provide working implementations of the proposed ideas.
This strengthens the case for its use in any of the participating community networks, or even to migrate between used data schemas.

Additionally, in the beginning of 2015, the \nodewatcher{} platform has been chosen to be the basis for the Commotion Wireless platform~\cite{Commotion_2015}.
In order to adapt to the specifics of the Commotion network, \nodewatcher{} has been extended to support security features, such as node-server mutual authentication and encryption of monitoring data, and support for data push in addition to data pull.
The platform's modular architecture has shown itself to be particularly suitable for customizing solutions for specific communities.

\subsection{Security, Availability and Federation}

Community networks are by their nature decentralized networks which grow in an ad-hoc fashion.
Some community networks may be concerned that having a centralized management system presents a single point of failure for the network, compromise its security or centralize the community too much.
In this section we analyse these concerns and argue that this may not be the case.

In order to support high availability scenarios, standard approaches like using multiple redundant servers and performing database replication, should be considered.
But even in case all redundancy fails, this will not affect the functioning of the actual community network, as operation of the nodes and routing protocols does not in any way rely on there being a \nodewatcher{} server.
Therefore, in the worst case, the only process that will be interrupted is network monitoring and support for managing nodes.

The issue of centralization can be addressed by federation. 
While \nodewatcher{} core does not explicitly support federated deployments, its modular nature enables communities to easily implement it as a module.
Assuming that there are multiple independent subcommunities within one larger community, there are two basic approaches in making \nodewatcher{} federated.

\begin{description}
    \item[Independent installations.] Each independent subcommunity or routing domain would have its own \nodewatcher{} installation which performs registration and monitoring for its own nodes.
    In this way, it would be completely independent from the centralized instance.
    If the community then wants to have an aggregated picture of the whole network, another top-level \nodewatcher{} instance may be deployed which will use pull and/or push from all the other subcommunity \nodewatcher{} installations.
    In this manner, the top-level installation would not support registration of nodes and would not monitor the nodes directly.
    Since in \nodewatcher{} these are all modules, they can be easily removed.
    Instead, the top-level instance would just get the data from subcommunity installations and use it as is.
    Because of the modular design, one would only need to develop a module that knows how to aggregate this data from multiple subcommunities and store it using the existing schema.

    \item[Single installation.] The problem with having multiple installations is that it may be hard to handle merge/split scenarios.
    So instead of having multiple installations, one could also use just a single \nodewatcher{} installation and just structure the nodes and permissions in such a way that each subcommunity has their space.
    This is similar to how the Guifi.net~\cite{Guifinode_2003,Vega_2012} dashboard splits nodes into zones.
    In this case, the server infrastructure would still be shared, but control would be distributed over multiple communities.
    Since there is no need for all the nodes to see each other (only \nodewatcher{} needs to be able to communicate with them), this is already possible using the current implementation.
    Currently, there is a module that supports \textit{projects}, but these have a completely flat structure.
    A flat structure does not scale to a large number of subcommunities.
    A better way would be to develop a module that would enable nicer visual grouping of nodes, using a parent-child concept similar to the Guifi.net zones.
    As far as topology and map visualizations go, the existing implementation already supports disconnected islands of nodes.
    Note that supporting large single installations (several thousand nodes and more) would most likely require performance optimizations in the monitoring modules, but since the whole monitoring pipeline is modular and designed to be easy to distribute over multiple servers, it can be improved upon by communities of such size.
\end{description}

The last concern regards security.
A centralized network management installation might be an attractive target for attackers.
Since \nodewatcher{} holds node configurations, those might contain sensitive information like passwords.
In order to minimize this exposure, public key authentication is supported and should be used instead of passwords whenever possible.
In this case only the public keys are stored by \nodewatcher{} and access to them does not grant access to the nodes themselves.
An additional security concern is for nodes to misreport data of other nodes, which would confuse the monitoring system, so it would display incorrect data.
This is why \nodewatcher{} also supports secure authentication of node data by using public keys mutually verified via the TLS~\cite{RFC_5246} protocol.

\section{Conclusion and Future Work}
\label{sec:conclusion}

In this paper we have presented \nodewatcher{}, a community network management platform, which is built around the core principle of modularity and extensibility, making it suitable for reuse by different community networks.
Devices are configured using platform-independent configuration which \nodewatcher{} can transform into deployable firmware images, eliminating any manual device configuration, reducing errors, and enabling participation of novice maintainers.
An embedded monitoring system enables live overview and validation of the whole community network.
We have shown how the system successfully operates in an actual community wireless network, \wlanslovenija{}, while it is also starting to be used by other network communities.

There are many possible improvements that could make various aspects of community networks easier to manage, e.g. help with radio signal and propagation planing using realistic models and geographic data, support for more routing protocols, and further optimizations for really large networks. Instead of developing all these new features ourselves, we envision the next step is to engage other network communities to develop features that are specific to their networks. We have already established partnerships with other community networks who are starting to work in this direction.

We also expect to see completely new features added to the platform by third-party developers.
For example, a warehouse module to help organize the inventory of equipment in community networks, a store module helping users order preconfigured devices, and local community and services modules to help people who are using the community networks to form better communities.
Using a strong and modular base is enabling further innovation and progress.

Data collected through \nodewatcher{} can be used in future research~\cite{Braem_2013}.
Researchers could analyze how community, wireless, or mesh networks operate over larger time spans.
Both on a technical level and also community and societal levels.
Researchers are already using the platform to monitor KORUZA~\cite{Mustafa_2013} deployments around the world and use it to further guide development and research in DIY free-space optics connectivity.

\section*{Acknowledgement}

The authors have been supported by the following institutions: Jernej Kos by the Slovenian Research Agency (Grant 1000-11-310153), by the Shuttleworth Foundation Flash Grant and by the NLnet Foundation (Grant 2014-05-015).
We would like to thank for their support everyone participating in the \wlanslovenija{} network and all the community networks, which responded to the survey.
Without this global community none of this work would be possible.
We would like to thank everyone who read early drafts of this paper, especially anonymous reviewers, and provided us with invaluable feedback.

\section*{References}
\bibliography{bibliography/references.bib}

\begin{thebibliography}{53}
\providecommand{\natexlab}[1]{#1}
\providecommand{\url}[1]{\texttt{#1}}
\providecommand{\urlprefix}{URL }
\expandafter\ifx\csname urlstyle\endcsname\relax
  \providecommand{\doi}[1]{doi:\discretionary{}{}{}#1}\else
  \providecommand{\doi}[1]{doi:\discretionary{}{}{}\begingroup
  \urlstyle{rm}\url{#1}\endgroup}\fi
\providecommand{\bibinfo}[2]{#2}

\bibitem[{Bruno et~al.(2005)Bruno, Conti, and Gregori}]{Bruno_2005}
\bibinfo{author}{R.~Bruno}, \bibinfo{author}{M.~Conti},
  \bibinfo{author}{E.~Gregori}, \bibinfo{title}{Mesh networks: commodity
  multihop ad hoc networks}, \bibinfo{journal}{Communications Magazine, IEEE}
  \bibinfo{volume}{43}~(\bibinfo{number}{3}) (\bibinfo{year}{2005})
  \bibinfo{pages}{123--131}, ISSN \bibinfo{issn}{0163-6804}.

\bibitem[{Frangoudis et~al.(2011)Frangoudis, Polyzos, and
  Kemerlis}]{Frangoudis_2011}
\bibinfo{author}{P.~Frangoudis}, \bibinfo{author}{G.~Polyzos},
  \bibinfo{author}{V.~Kemerlis}, \bibinfo{title}{Wireless community networks:
  an alternative approach for nomadic broadband network access},
  \bibinfo{journal}{Communications Magazine, IEEE}
  \bibinfo{volume}{49}~(\bibinfo{number}{5}) (\bibinfo{year}{2011})
  \bibinfo{pages}{206--213}, ISSN \bibinfo{issn}{0163-6804}.

\bibitem[{Akyildiz et~al.(2005)Akyildiz, Wang, and Wang}]{Akyildiz_2005}
\bibinfo{author}{I.~F. Akyildiz}, \bibinfo{author}{X.~Wang},
  \bibinfo{author}{W.~Wang}, \bibinfo{title}{Wireless mesh networks: a survey},
  \bibinfo{journal}{Computer Networks}
  \bibinfo{volume}{47}~(\bibinfo{number}{4}) (\bibinfo{year}{2005})
  \bibinfo{pages}{445 -- 487}, ISSN \bibinfo{issn}{1389-1286}.

\bibitem[{Mustafa and Thomsen(2013)}]{Mustafa_2013}
\bibinfo{author}{L.~Mustafa}, \bibinfo{author}{B.~Thomsen},
  \bibinfo{title}{Reintroducing free-space optical technology to community
  wireless networks}, in: \bibinfo{booktitle}{19th Americas Conference on
  Information Systems, Chicago}, \bibinfo{year}{2013}.

\bibitem[{{Open Technology Institute}(2013)}]{RedHook_2013}
\bibinfo{author}{{Open Technology Institute}}, \bibinfo{title}{Case Study: Red
  Hook Initiative WiFi \& Tidepools},
  \urlprefix\url{http://oti.newamerica.net/blogposts/2013/case_study_red_hook_initiative_wifi_tidepools-78575},
  \bibinfo{year}{2013}.

\bibitem[{AWM(2002)}]{AWMN}
\bibinfo{title}{Athens Wireless Metropolitan Network},
  \urlprefix\url{https://awmn.net}, \bibinfo{year}{2002}.

\bibitem[{wla(2009)}]{wlanslovenija_2009}
\bibinfo{title}{wlan slovenija}, \urlprefix\url{http://wlan-si.net},
  \bibinfo{year}{2009}.

\bibitem[{gui(2003)}]{guifi_2003}
\bibinfo{title}{Guifi.net}, \urlprefix\url{http://guifi.net},
  \bibinfo{year}{2003}.

\bibitem[{Fun(2003)}]{Funkfeuer_2003}
\bibinfo{title}{Funkfeuer}, \urlprefix\url{http://www.funkfeuer.at},
  \bibinfo{year}{2003}.

\bibitem[{Fre(2003)}]{Freifunk_2003}
\bibinfo{title}{Freifunk}, \urlprefix\url{http://freifunk.net},
  \bibinfo{year}{2003}.

\bibitem[{Butler(2013)}]{WNDW_2013}
\bibinfo{editor}{J.~S. Butler} (Ed.), \bibinfo{title}{Wireless Networking in
  the Developing World}, \bibinfo{publisher}{CreateSpace Independent Publishing
  Platform}, ISBN \bibinfo{isbn}{978-1484039359}, \bibinfo{year}{2013}.

\bibitem[{int(2010)}]{interop_2010}
\bibinfo{title}{Open Networks Interoperability},
  \urlprefix\url{http://interop.wlan-si.net}, \bibinfo{year}{2010}.

\bibitem[{cnm(2007)}]{cnml_2007}
\bibinfo{title}{Community Network Mark Up Language Project},
  \urlprefix\url{http://cnml.info}, \bibinfo{year}{2007}.

\bibitem[{Murray et~al.(2010)Murray, Dixon, and Koziniec}]{Murray_2010}
\bibinfo{author}{D.~Murray}, \bibinfo{author}{M.~Dixon},
  \bibinfo{author}{T.~Koziniec}, \bibinfo{title}{An experimental comparison of
  routing protocols in multi hop ad hoc networks}, in:
  \bibinfo{booktitle}{Telecommunication Networks and Applications Conference
  (ATNAC), 2010 Australasian}, \bibinfo{pages}{159--164}, \bibinfo{year}{2010}.

\bibitem[{Neumann et~al.(2012)Neumann, Lopez, and Navarro}]{Neumann_2012}
\bibinfo{author}{A.~Neumann}, \bibinfo{author}{E.~Lopez},
  \bibinfo{author}{L.~Navarro}, \bibinfo{title}{An evaluation of BMX6 for
  community wireless networks}, in: \bibinfo{booktitle}{Wireless and Mobile
  Computing, Networking and Communications (WiMob), 2012 IEEE 8th International
  Conference on}, ISSN \bibinfo{issn}{2160-4886}, \bibinfo{pages}{651--658},
  \bibinfo{year}{2012}.

\bibitem[{Neumann et~al.(2013)Neumann, Navarro, Baig, and
  Escrich}]{Neumann_2013}
\bibinfo{author}{A.~Neumann}, \bibinfo{author}{L.~Navarro},
  \bibinfo{author}{R.~Baig}, \bibinfo{author}{P.~Escrich},
  \bibinfo{title}{Receiver-driven routing for community mesh networks}, in:
  \bibinfo{booktitle}{World of Wireless, Mobile and Multimedia Networks
  (WoWMoM), 2013 IEEE 14th International Symposium and Workshops on a},
  \bibinfo{pages}{1--7}, \doi{\bibinfo{doi}{10.1109/WoWMoM.2013.6583481}},
  \bibinfo{year}{2013}.

\bibitem[{Siddiqui and Hong(2007)}]{Siddiqui_2007}
\bibinfo{author}{M.~Siddiqui}, \bibinfo{author}{C.~S. Hong},
  \bibinfo{title}{Security Issues in Wireless Mesh Networks}, in:
  \bibinfo{booktitle}{Multimedia and Ubiquitous Engineering, 2007. MUE '07.
  International Conference on}, \bibinfo{pages}{717--722},
  \bibinfo{year}{2007}.

\bibitem[{Vega et~al.(2012)Vega, Cerda-Alabern, Navarro, and
  Meseguer}]{Vega_2012}
\bibinfo{author}{D.~Vega}, \bibinfo{author}{L.~Cerda-Alabern},
  \bibinfo{author}{L.~Navarro}, \bibinfo{author}{R.~Meseguer},
  \bibinfo{title}{Topology patterns of a community network: Guifi.net}, in:
  \bibinfo{booktitle}{Wireless and Mobile Computing, Networking and
  Communications (WiMob), 2012 IEEE 8th International Conference on},
  \bibinfo{organization}{IEEE}, \bibinfo{pages}{612--619},
  \bibinfo{year}{2012}.

\bibitem[{Zakrzewska et~al.(2008)Zakrzewska, Koszalka, and
  Pozniak-Koszalka}]{Zakrzewska_2008}
\bibinfo{author}{A.~Zakrzewska}, \bibinfo{author}{L.~Koszalka},
  \bibinfo{author}{I.~Pozniak-Koszalka}, \bibinfo{title}{Performance Study of
  Routing Protocols for Wireless Mesh Networks}, in:
  \bibinfo{booktitle}{Systems Engineering, 2008. ICSENG '08. 19th International
  Conference on}, \bibinfo{pages}{331--336}, \bibinfo{year}{2008}.

\bibitem[{Braem et~al.(2014)Braem, Bergs, Avonts, and Blondia}]{Braem_2014}
\bibinfo{author}{B.~Braem}, \bibinfo{author}{J.~Bergs},
  \bibinfo{author}{J.~Avonts}, \bibinfo{author}{C.~Blondia},
  \bibinfo{title}{Mapping a community network}, in: \bibinfo{booktitle}{Global
  Information Infrastructure and Networking Symposium (GIIS), 2014},
  \bibinfo{pages}{1--8}, \doi{\bibinfo{doi}{10.1109/GIIS.2014.6934252}},
  \bibinfo{year}{2014}.

\bibitem[{Cerd\`{a}-Alabern et~al.(2013)Cerd\`{a}-Alabern, Neumann, and
  Escrich}]{Cerda_2013}
\bibinfo{author}{L.~Cerd\`{a}-Alabern}, \bibinfo{author}{A.~Neumann},
  \bibinfo{author}{P.~Escrich}, \bibinfo{title}{Experimental Evaluation of a
  Wireless Community Mesh Network}, in: \bibinfo{booktitle}{Proceedings of the
  16th ACM International Conference on Modeling, Analysis \&\#38; Simulation of
  Wireless and Mobile Systems}, MSWiM '13, \bibinfo{publisher}{ACM},
  \bibinfo{address}{New York, NY, USA}, ISBN \bibinfo{isbn}{978-1-4503-2353-6},
  \bibinfo{pages}{23--30}, \doi{\bibinfo{doi}{10.1145/2507924.2507960}},
  \bibinfo{year}{2013}.

\bibitem[{Maccari and Cigno(2015)}]{Maccari_2015}
\bibinfo{author}{L.~Maccari}, \bibinfo{author}{R.~L. Cigno}, \bibinfo{title}{A
  week in the life of three large Wireless Community Networks},
  \bibinfo{journal}{Ad Hoc Networks} \bibinfo{volume}{24, Part B}
  (\bibinfo{year}{2015}) \bibinfo{pages}{175 -- 190}, ISSN
  \bibinfo{issn}{1570-8705},
  \doi{\bibinfo{doi}{http://dx.doi.org/10.1016/j.adhoc.2014.07.016}},
  \bibinfo{note}{modeling and Performance Evaluation of Wireless Ad-Hoc
  Networks}.

\bibitem[{Cac(2004)}]{Cacti_2004}
\bibinfo{title}{Cacti}, \urlprefix\url{http://www.cacti.net},
  \bibinfo{year}{2004}.

\bibitem[{Nag(1999)}]{Nagios_1999}
\bibinfo{title}{Nagios}, \urlprefix\url{http://www.nagios.org},
  \bibinfo{year}{1999}.

\bibitem[{Oetiker(2001)}]{SmokePing_2001}
\bibinfo{author}{T.~Oetiker}, \bibinfo{title}{SmokePing},
  \urlprefix\url{http://oss.oetiker.ch/smokeping/}, \bibinfo{year}{2001}.

\bibitem[{Zab(2004)}]{Zabbix_2004}
\bibinfo{title}{Zabbix}, \urlprefix\url{http://www.zabbix.com},
  \bibinfo{year}{2004}.

\bibitem[{Pup(2005)}]{Puppet_2005}
\bibinfo{title}{Puppet}, \urlprefix\url{http://puppetlabs.com},
  \bibinfo{year}{2005}.

\bibitem[{Sal(2011)}]{Salt_2011}
\bibinfo{title}{SaltStack}, \urlprefix\url{http://saltstack.com},
  \bibinfo{year}{2011}.

\bibitem[{{Guifi Community}(2003)}]{Guifinode_2003}
\bibinfo{author}{{Guifi Community}}, \bibinfo{title}{Guifi.net Node Database},
  \urlprefix\url{https://guifi.net/en/guifi_zones}, \bibinfo{year}{2003}.

\bibitem[{{Athens Wireless Metropolitan Network}(2002)}]{AWMN_WIND_2002}
\bibinfo{author}{{Athens Wireless Metropolitan Network}},
  \bibinfo{title}{WiND}, \urlprefix\url{https://wind.awmn.net},
  \bibinfo{year}{2002}.

\bibitem[{O'Neil(2008)}]{ONeil_2008}
\bibinfo{author}{E.~J. O'Neil}, \bibinfo{title}{Object/Relational Mapping 2008:
  Hibernate and the Entity Data Model (Edm)}, in:
  \bibinfo{booktitle}{Proceedings of the 2008 ACM SIGMOD International
  Conference on Management of Data}, SIGMOD '08, \bibinfo{publisher}{ACM},
  \bibinfo{address}{New York, NY, USA}, ISBN \bibinfo{isbn}{978-1-60558-102-6},
  \bibinfo{pages}{1351--1356}, \bibinfo{year}{2008}.

\bibitem[{Capoano et~al.(2012)}]{Nodeshot_2012}
\bibinfo{author}{F.~Capoano}, et~al., \bibinfo{title}{Nodeshot},
  \urlprefix\url{https://github.com/ninuxorg/nodeshot}, \bibinfo{year}{2012}.

\bibitem[{Tanzer et~al.(2012)}]{Funkfeuer_2012}
\bibinfo{author}{C.~Tanzer}, et~al., \bibinfo{title}{Funkfeuer NodeDB},
  \urlprefix\url{https://github.com/FFM}, \bibinfo{year}{2012}.

\bibitem[{Bernstein and Melnik(2007)}]{Bernstein_2007}
\bibinfo{author}{P.~A. Bernstein}, \bibinfo{author}{S.~Melnik},
  \bibinfo{title}{Model Management 2.0: Manipulating Richer Mappings}, in:
  \bibinfo{booktitle}{Proceedings of the 2007 ACM SIGMOD International
  Conference on Management of Data}, SIGMOD '07, \bibinfo{publisher}{ACM},
  \bibinfo{address}{New York, NY, USA}, ISBN \bibinfo{isbn}{978-1-59593-686-8},
  \bibinfo{pages}{1--12}, \bibinfo{year}{2007}.

\bibitem[{dja(2005)}]{django_2005}
\bibinfo{title}{Django: The web framework for perfectionists with deadlines},
  \urlprefix\url{https://www.djangoproject.com/}, \bibinfo{year}{2005}.

\bibitem[{Ove(2013)}]{Overextend_2013}
\bibinfo{title}{django-overextends},
  \urlprefix\url{https://github.com/wlanslovenija/django-overextends/tree/wlanslovenija},
  \bibinfo{year}{2013}.

\bibitem[{Peterson and Norman(1977)}]{Peterson_1977}
\bibinfo{author}{J.~L. Peterson}, \bibinfo{author}{T.~A. Norman},
  \bibinfo{title}{Buddy Systems}, \bibinfo{journal}{Commun. ACM}
  \bibinfo{volume}{20}~(\bibinfo{number}{6}) (\bibinfo{year}{1977})
  \bibinfo{pages}{421--431}, ISSN \bibinfo{issn}{0001-0782}.

\bibitem[{Ope(2004)}]{OpenWrt_2004}
\bibinfo{title}{OpenWrt}, \urlprefix\url{https://openwrt.org},
  \bibinfo{year}{2004}.

\bibitem[{Rou(1995)}]{RouterOS_1995}
\bibinfo{title}{RouterOS}, \urlprefix\url{http://www.mikrotik.com/software},
  \bibinfo{year}{1995}.

\bibitem[{Hykes(2013)}]{Docker_2013}
\bibinfo{author}{S.~Hykes}, \bibinfo{title}{Docker},
  \urlprefix\url{https://www.docker.com}, \bibinfo{year}{2013}.

\bibitem[{Oetiker(1999)}]{Oetiker_1999}
\bibinfo{author}{T.~Oetiker}, \bibinfo{title}{Round-robin Database Tool},
  \urlprefix\url{http://oss.oetiker.ch/rrdtool}, \bibinfo{year}{1999}.

\bibitem[{Tok(2007)}]{TokuMX_2007}
\bibinfo{title}{TokuMX},
  \urlprefix\url{http://www.tokutek.com/tokumx-for-mongodb},
  \bibinfo{year}{2007}.

\bibitem[{Mon(2007)}]{MongoDB_2007}
\bibinfo{title}{MongoDB}, \urlprefix\url{http://www.mongodb.org},
  \bibinfo{year}{2007}.

\bibitem[{Brodal and Fagerberg(2003)}]{Brodal_2003}
\bibinfo{author}{G.~S. Brodal}, \bibinfo{author}{R.~Fagerberg},
  \bibinfo{title}{Lower Bounds for External Memory Dictionaries}, in:
  \bibinfo{booktitle}{Proceedings of the Fourteenth Annual ACM-SIAM Symposium
  on Discrete Algorithms}, SODA '03, \bibinfo{publisher}{Society for Industrial
  and Applied Mathematics}, \bibinfo{address}{Philadelphia, PA, USA}, ISBN
  \bibinfo{isbn}{0-89871-538-5}, \bibinfo{pages}{546--554},
  \bibinfo{year}{2003}.

\bibitem[{Bender et~al.(2007)Bender, Farach-Colton, Fineman, Fogel, Kuszmaul,
  and Nelson}]{Bender_2007}
\bibinfo{author}{M.~A. Bender}, \bibinfo{author}{M.~Farach-Colton},
  \bibinfo{author}{J.~T. Fineman}, \bibinfo{author}{Y.~R. Fogel},
  \bibinfo{author}{B.~C. Kuszmaul}, \bibinfo{author}{J.~Nelson},
  \bibinfo{title}{Cache-oblivious Streaming B-trees}, in:
  \bibinfo{booktitle}{Proceedings of the Nineteenth Annual ACM Symposium on
  Parallel Algorithms and Architectures}, SPAA '07, \bibinfo{publisher}{ACM},
  \bibinfo{address}{New York, NY, USA}, ISBN \bibinfo{isbn}{978-1-59593-667-7},
  \bibinfo{pages}{81--92}, \bibinfo{year}{2007}.

\bibitem[{Dat(2012)}]{Datastream_2012}
\bibinfo{title}{Datastream Library},
  \urlprefix\url{https://github.com/wlanslovenija/datastream},
  \bibinfo{year}{2012}.

\bibitem[{łgorzata Steinder and Sethi(2004)}]{Steinder_2004}
\bibinfo{author}{M.~łgorzata Steinder}, \bibinfo{author}{A.~S. Sethi},
  \bibinfo{title}{A survey of fault localization techniques in computer
  networks}, \bibinfo{journal}{Science of Computer Programming}
  \bibinfo{volume}{53}~(\bibinfo{number}{2}) (\bibinfo{year}{2004})
  \bibinfo{pages}{165 -- 194}, ISSN \bibinfo{issn}{0167-6423},
  \bibinfo{note}{topics in System Administration}.

\bibitem[{Tra(2003)}]{Trac_2003}
\bibinfo{title}{The Trac Project}, \urlprefix\url{http://trac.edgewall.org/},
  \bibinfo{year}{2003}.

\bibitem[{Tra(2004)}]{TracHacks_2004}
\bibinfo{title}{Trac Hacks}, \urlprefix\url{http://trac-hacks.org/},
  \bibinfo{year}{2004}.

\bibitem[{Dja(2010)}]{DjangoPackages_2010}
\bibinfo{title}{Django Packages},
  \urlprefix\url{https://www.djangopackages.com/}, \bibinfo{year}{2010}.

\bibitem[{King(2015)}]{Commotion_2015}
\bibinfo{author}{J.~King}, \bibinfo{title}{Announcing collaboration with wlan
  slovenija},
  \urlprefix\url{https://commotionwireless.net/blog/2015/04/22/announcing-commotion-collaboration-with-wlanslovenija},
  \bibinfo{year}{2015}.

\bibitem[{Dierks and Rescorla(2008)}]{RFC_5246}
\bibinfo{author}{T.~Dierks}, \bibinfo{author}{E.~Rescorla},
  \bibinfo{title}{{The Transport Layer Security (TLS) Protocol Version 1.2}},
  \bibinfo{howpublished}{RFC 5246 (Proposed Standard)},
  \urlprefix\url{http://www.ietf.org/rfc/rfc5246.txt}, \bibinfo{note}{updated
  by RFCs 5746, 5878, 6176, 7465, 7507}, \bibinfo{year}{2008}.

\bibitem[{Braem et~al.(2013)Braem, Blondia, Barz, Rogge, Freitag, Navarro,
  Bonicioli, Papathanasiou, Escrich, Baig Vi\~{n}as, Kaplan, Neumann, Vilata~i
  Balaguer, Tatum, and Matson}]{Braem_2013}
\bibinfo{author}{B.~Braem}, \bibinfo{author}{C.~Blondia},
  \bibinfo{author}{C.~Barz}, \bibinfo{author}{H.~Rogge},
  \bibinfo{author}{F.~Freitag}, \bibinfo{author}{L.~Navarro},
  \bibinfo{author}{J.~Bonicioli}, \bibinfo{author}{S.~Papathanasiou},
  \bibinfo{author}{P.~Escrich}, \bibinfo{author}{R.~Baig Vi\~{n}as},
  \bibinfo{author}{A.~L. Kaplan}, \bibinfo{author}{A.~Neumann},
  \bibinfo{author}{I.~Vilata~i Balaguer}, \bibinfo{author}{B.~Tatum},
  \bibinfo{author}{M.~Matson}, \bibinfo{title}{A Case for Research with and on
  Community Networks}, \bibinfo{journal}{SIGCOMM Comput. Commun. Rev.}
  \bibinfo{volume}{43}~(\bibinfo{number}{3}) (\bibinfo{year}{2013})
  \bibinfo{pages}{68--73}, ISSN \bibinfo{issn}{0146-4833}.

\end{thebibliography}

\end{document}